\documentclass[twocolumn,prd,amssymb,aps,nofootinbib,showpacs]{revtex4}
\usepackage[usenames,dvipsnames]{color}
\usepackage{ulem}
\usepackage{graphicx}
\usepackage{amssymb}
\usepackage{amsmath}
\usepackage{epstopdf}

\begin{document}
\title{Extracting equation of state parameters from black hole-neutron star mergers: aligned-spin black holes and a preliminary waveform model}
\author{Benjamin D. Lackey$^1$, Koutarou Kyutoku$^2$, Masaru Shibata$^3$, Patrick R. Brady$^4$, John L. Friedman$^4$}

\affiliation{
$^1$Department of Physics, Princeton University, Princeton, NJ 08544, USA\\
$^2$Theory Center, Institute of Particles and Nuclear Studies, KEK, Tsukuba, 305-0801, Japan\\
$^3$Yukawa Institute for Theoretical Physics, Kyoto University, Kyoto 606-8502, Japan\\
$^4$Department of Physics, University of Wisconsin--Milwaukee, Milwaukee, WI 53201, USA
}

\begin{abstract}
Information about the neutron-star equation of state is encoded in the waveform of a black hole-neutron star system through tidal interactions and the possible tidal disruption of the neutron star. During the inspiral this information depends on the tidal deformability $\Lambda$ of the neutron star, and we find that $\Lambda$ is the best measured parameter during the merger and ringdown as well. We performed 134 simulations where we systematically varied the equation of state as well as the mass ratio, neutron star mass, and aligned spin of the black hole. Using these simulations we have developed an analytic representation of the full inspiral-merger-ringdown waveform calibrated to these numerical waveforms, and we use this analytic waveform to estimate the accuracy to which $\Lambda$ can be measured with gravitational-wave detectors. We find that although the inspiral tidal signal is small, \textit{coherently} combining this signal with the merger-ringdown matter effect improves the measurability of $\Lambda$ by a factor of $\sim 3$ over using just the merger-ringdown matter effect alone. However, incorporating correlations between all the waveform parameters then decreases the measurability of $\Lambda$ by a factor of $\sim 3$. The uncertainty in $\Lambda$ increases with the mass ratio, but decreases as the black hole spin increases. Overall, a single Advanced LIGO detector can measure $\Lambda$ for mass ratios $Q = 2$--5, black hole spins $J_{\rm BH}/M_{\rm BH}^2 = -0.5$--0.75, neutron star masses $M_{\rm NS} = 1.2M_\odot$--$1.45M_\odot$, and an optimally oriented distance of 100~Mpc to a 1-$\sigma$ uncertainty of $\sim 10$\%--100\%. For the proposed Einstein Telescope, the uncertainty in $\Lambda$ is an order of magnitude smaller. 

\end{abstract}


\maketitle

\section{Introduction}

By the end of the decade a network of second generation gravitational-wave detectors, including the two Advanced LIGO (aLIGO) detectors~\cite{Harry2010}, Advanced Virgo~\cite{Acernese2009}, \mbox{KAGRA}~\cite{Somiya2012} (formerly LCGT), and possibly LIGO-India~\cite{IyerSouradeepUnnikrishnan2011}, will likely be making routine detections. Future ground based detectors such as the third generation Einstein Telescope (ET)~\cite{Punturo2010}, with an order of magnitude higher sensitivity, are also in the planning stages, and may be operational in the next decade. A primary goal of these detectors is extracting from the gravitational waveform information about the sources. Of particular interest are compact binaries whose waveform encodes the sky location, orientation, distance, masses, spins, and for compact binaries containing neutron stars (NS), information about the neutron-star equation of state (EOS).

The study of EOS effects during binary inspiral has focused mainly on binary neutron star (BNS) systems. Work by~\cite{Kochanek1992, LaiRasioShapiro1994, MoraWill2004, BertiIyerWill2008} showed that EOS information could be imprinted in the gravitational waveform through tidal interactions. In the adiabatic approximation, the quadrupole moment $Q_{ij}$ of one star depends on the tidal field $\mathcal{E}_{ij}$ from the monopole of the other star through the relation $Q_{ij} = -\lambda \mathcal{E}_{ij}$, where $\lambda$ is the EOS dependent tidal deformability and is related to the neutron star's dimensionless Love number $k_2$ and radius $R$ through the relation $\lambda = \frac{2}{3G} k_2 R^5$. The leading ($\ell = 2$) relativistic tidal Love number $k_2$ was first calculated in Ref.~\cite{Hinderer2008} for polytropic EOS, then for EOS with hadronic and quark matter~\cite{HindererLackeyLangRead2010, PostnikovPrakashLattimer2010}, as well as for EOS with analytic solutions to the stellar structure equations~\cite{PostnikovPrakashLattimer2010}. Its effect on the binary inspiral (including the contribution due to tidally excited f-modes) was calculated to leading order~\cite{FlanaganHinderer2008}, and later extended to 1PN order~\cite{VinesFlanagan2010, VinesFlanaganHinderer2011}. The gravitoelectric and gravitomagnetic tidal Love numbers for higher multipoles were calculated in~\cite{DamourNagar2009tidal, BinningtonPoisson2009}. The energy has now been calculated to 2PN order in the tidal corrections in the effective one body (EOB) formalism, including $\ell = 2$ and 3 gravitoelectic interactions and the $\ell = 2$ gravitomagnetic interaction, using the effective action approach~\cite{BiniDamourFaye2012}, and most terms in the EOB waveform are now known to 2.5PN order in the tidal interactions~\cite{DamourNagarVillain2012}. Finally, the accuracy of the adiabatic approximation to tidal interactions was calculated using an affine model, and a Love function was found that corrects for this approximation and asymptotically approaches the Love number for large binary separations~\cite{FerrariGualtieriMaselli2012, MaselliGualtieriPannarale2012}.

The measurability of tidal parameters by detectors with the sensitivity of aLIGO and ET was examined for BNS inspiral for gravitational wave frequencies below 450Hz~\cite{FlanaganHinderer2008} using polytropic EOS as well as for theoretical hadronic and quark matter EOS~\cite{HindererLackeyLangRead2010}.  The studies found that tidal interactions were observable during this early inspiral stage (prior to the last $\sim 20$ gravitational wave cycles before merger) only for stiff EOS and NS masses below 1.4~$M_\odot$. On the other hand, using tidal corrections up to 2.5PN order in the EOB approach, it was found that tidal parameters are in fact observable when including the extra $\sim 20$ gravitational wave cycles up to the point of contact~\cite{DamourNagarVillain2012}.

Numerical simulations have also been used to study the measurability of matter effects during the late inspiral of BNS systems. Read et al.~\cite{ReadMarkakisShibata2009} examined the measurability of EOS information during the last few orbits, using numerical simulations assuming that non EOS parameters do not correlate significantly with EOS parameters. They found that the NS radius could be measured, using only the last few orbits, to an accuracy of $\sim10$\%. In addition, using simulations in the conformally flat approximation, Ref.~\cite{BausweinJankaHebeler2012} found that the NS radius could be determined to a comparable accuracy from the post-merger phase.

Numerical work is also in progress to verify the accuracy of the inspiral tidal description using BNS simulations. Initial studies comparing quasiequilibrium sequences~\cite{DamourNagar2010} and full hydrodynamic simulations~\cite{BaiottiDamour2010, BaiottiDamour2011} with post-Newtonian and EOB tidal descriptions found noticeable differences, suggesting large corrections to the analytic description might be necessary. However, later comparisons have highlighted the importance of the hydrodynamics treatment, numerical resolution, and waveform extraction radius in determining the tidal contribution to the waveform with numerical simulations~\cite{BernuzziNagarThierfelder2012, HotokazakaKyutokuShabata2013}. These later comparisons found that within numerical error the improved simulations are consistent with the 2PN accurate EOB tidal description up to merger~\cite{BernuzziNagarThierfelder2012}, or at least up to the last $\sim 10$ gravitational wave cycles~\cite{HotokazakaKyutokuShabata2013}.

Work is now underway to understand the measurability of EOS information in black hole-neutron star (BHNS) systems as well. Studies using the inspiral waveform with tidal corrections up to 1PN order found that BHNS waveforms are not distinguishable from binary black hole (BBH) waveforms before the end of inspiral when the frequency reaches the ISCO frequency or tidal disruption frequency~\cite{PannaraleRezzollaOhmeRead2011}. This conclusion was also found to be true for BHNS systems with spinning black holes where this ending frequency may be larger than for a nonspinning BH. On the other hand, work examining a possible cutoff in the gravitational wave amplitude due to tidal disruption of the neutron star by the black hole suggested that the NS radius may, in fact, be measurable with second generation detectors~\cite{Vallisneri2000, FerrariGualtieriPannarale2009, FerrariGualtieriPannarale2010}. 

Several numerical simulations have also been carried out to examine EOS dependent effects during the end of the BHNS inspiral~\cite{ShibataKyutokuYamamotoTaniguchi2009, KyutokuShibataTaniguchi2010, Duez2010, KyutokuOkawaShibataTaniguchi2011}, and these simulations now include BH to NS mass ratios of up to $Q = M_{\rm BH}/M_{\rm NS} = 7.1$ and black hole spins of up to $\chi_{\rm BH} = J_{\rm BH}/M_{\rm BH}^2 = 0.9$, where $J_{\rm BH}$ is the black hole's angular momentum~\cite{FoucartDuezKidder2012, FoucartDeatonDuez2012}. In Ref.~\cite{LackeyKyutoku2012} (hereafter Paper~I), we examined numerical simulations of the last few orbits, merger, and ringdown for systems with nonspinning black holes and low mass ratios of $Q = 2$ and 3. We found that when considering only the merger and ringdown, the tidal deformability $\Lambda$ was the best measured EOS parameter and was marginally measurable for second generation detectors. As in Paper~I, we define a dimensionless version $\Lambda$ of the the tidal deformability using the NS mass
\begin{equation}
\Lambda := G\lambda \left(\frac{c^2}{GM_{\rm NS}}\right)^5 
= \frac{2}{3} k_2 \left(\frac{c^2R}{GM_{\rm NS}}\right)^5. 
\end{equation}

In this paper we repeat the analysis of Paper~I for mass ratios up to $Q = 5$ and black hole spins from $\chi_{\rm BH} = -0.5$ to 0.75. We will also address many of the simplifications used in Paper~I that can have a significant impact on the detectability of EOS parameters. Previously, we considered only the tidal information that could be obtained from the merger and ringdown, ignoring the small accumulating phase drift during the inspiral that results from tidal interactions. We will find that \textit{coherently} adding the slow tidal phase drift from the inspiral to the tidal effect during the merger and ringdown (as was also discussed in Ref.~\cite{FoucartDeatonDuez2012}) can improve the measurability of tidal parameters by a factor of $\sim 3$ over just the merger and ringdown. 

In Paper~I we also ignored possible correlations between the tidal parameter $\Lambda$ and the other binary parameters when estimating the measurability of $\Lambda$ with the Fisher matrix approximation. We have addressed this issue by developing a frequency-domain, analytic BHNS waveform model fit to our BHNS simulations, and based on analytic BBH waveform models. This fit allows us to accurately evaluate derivatives in the Fisher matrix and evaluate correlations between the tidal parameter $\Lambda$ and the other parameters. We find that although these correlations are not nearly as strong as with other pairs of parameters, they can increase the uncertainty in $\Lambda$ by a factor of $\sim 3$. Overall, we find that the measurability of $\Lambda$ using the improved methods in this paper are about the same as the estimates in Paper~I, where only the merger and ringdown were considered and correlations with the other parameters were presumed to be negligible. Finally, we calculate the systematic error in our BHNS waveform model and find that it is significantly smaller than the statistical error for aLIGO and comparable to the statistical error for the ET detector.

In Sec.~\ref{sec:2}, we briefly describe the EOS and numerical methods used to generate our BHNS waveforms, as well as the waveform behavior as a function of mass ratio, BH spin, and EOS. We then construct hybrid waveforms that join the numerical waveforms to inspiral waveforms in Sec.~\ref{sec:3}. We then provide an overview of parameter estimation and show that $\Lambda$ is the best measured EOS parameter in Secs.~\ref{sec:4} and~\ref{sec:5}. Finally, to understand correlations between the parameters, we construct a phenomenological BHNS waveform fit to our hybrid BHNS waveforms in Sec.~\ref{sec:6}, then estimate the ability to measure $\Lambda$ with aLIGO and ET in Sec.~\ref{sec:7}. We conclude in Sec.~\ref{sec:8} with a brief discussion of some of the improvements needed to generate templates  accurate enough to use in data analysis pipelines.

\textit{Conventions}: We use the following sign convention for the Fourier transform of a signal $x(t)$
\begin{equation}
\tilde x(f)=\int_{-\infty}^\infty x(t) e^{+2 \pi i f t}\,dt,
\end{equation}
and we will decompose the complex Fourier transform into amplitude and phase as $\tilde h(f) = |\tilde h(f)| e^{+i\Phi(f)}$. These sign conventions are opposite those of Paper~I, and are chosen to agree with those of the PhenomC waveform model~\cite{Santamaria2010}, which we use extensively to construct hybrid waveforms. In addition, we set $G = c = 1$ unless otherwise stated.

\section{Simulations}
\label{sec:2}

Following Paper~I on nonspinning BHNS systems, we perform a large set of simulations where we systematically vary the parameters of a parametrized EOS, then look for the combination of parameters that are best extracted from gravitational-wave observations. Specifically we choose a simplified two-parameter version of the piecewise polytrope EOS introduced in Ref.~\cite{ReadLackey2009}. For this EOS, the pressure $p$ in the rest-mass density interval $\rho_{i-1} < \rho < \rho_i$ is
\begin{equation}
p(\rho) = K_i \rho^{\Gamma_i},
\end{equation}
where $K_i$ is a constant, and $\Gamma_i$ is the adiabatic index. We fix the crust EOS defined by densities below the transition density $\rho_0$. In the crust, $K_0 = 3.5966 \times 10^{13}$ in cgs units and $\Gamma_0 = 1.3569$, such that the pressure at $10^{13}$~g/cm$^3$ is $1.5689 \times 10^{31}$~dyne/cm$^2$. Above the transition density $\rho_0$, the core EOS is parametrized by the two parameters $p_1$ and $\Gamma_1$. The pressure $p_1$ is defined as the pressure at $\rho_1 = 10^{14.7}$~g/cm$^3$ and the adiabatic index $\Gamma_1$ of the core will, for simplicity, be written $\Gamma$. The constant $K_1$ for the core is then given by $K_1 = p_1/\rho_1^{\Gamma}$. Once the two parameters $p_1$ and $\Gamma$ are set, the dividing density $\rho_0$ between the crust and the core is given by the density where the crust and core EOS intersect: $\rho_0 = (K_0/K_1)^{1/(\Gamma - \Gamma_0)}$. Finally, given this EOS, the energy density $\epsilon$ can be evaluated by integrating the first law of thermodynamics
\begin{equation}
d\frac{\epsilon}{\rho} = - p d\frac{1}{\rho}.
\end{equation}

As discussed in more detail in Paper~I, quasiequilibrium configurations are used as initial data for the simulations~\cite{KyutokuShibataTaniguchi2009, KyutokuOkawaShibataTaniguchi2011}, and are computed using the spectral-method library \texttt{LORENE}~\cite{LORENE}. The numerical simulations are performed using the adaptive-mesh refinement code \texttt{SACRA}~\cite{YamamotoShibataTaniguchi2008}. To obtain the gravitational waveform $h_+ - i h_\times$, the outgoing part of the Weyl scalar $\Psi_4 = \ddot h_+ - i \ddot h_\times$ is extracted from these simulations at a finite coordinate radius, and is then integrated twice using a method known as Fixed Frequency Integration~\cite{ReisswigPollney2010}. Specifically, we take the Fourier transform of $\Psi_4$, then integrate twice in time by dividing by $(2 \pi i f)^2$. Low frequency components are filtered out as in Paper~I, and the inverse Fourier transform is then taken to find $h_+ - i h_\times$ in the time domain.

\begin{figure}[!htb]
\begin{center}
\includegraphics[width=80mm]{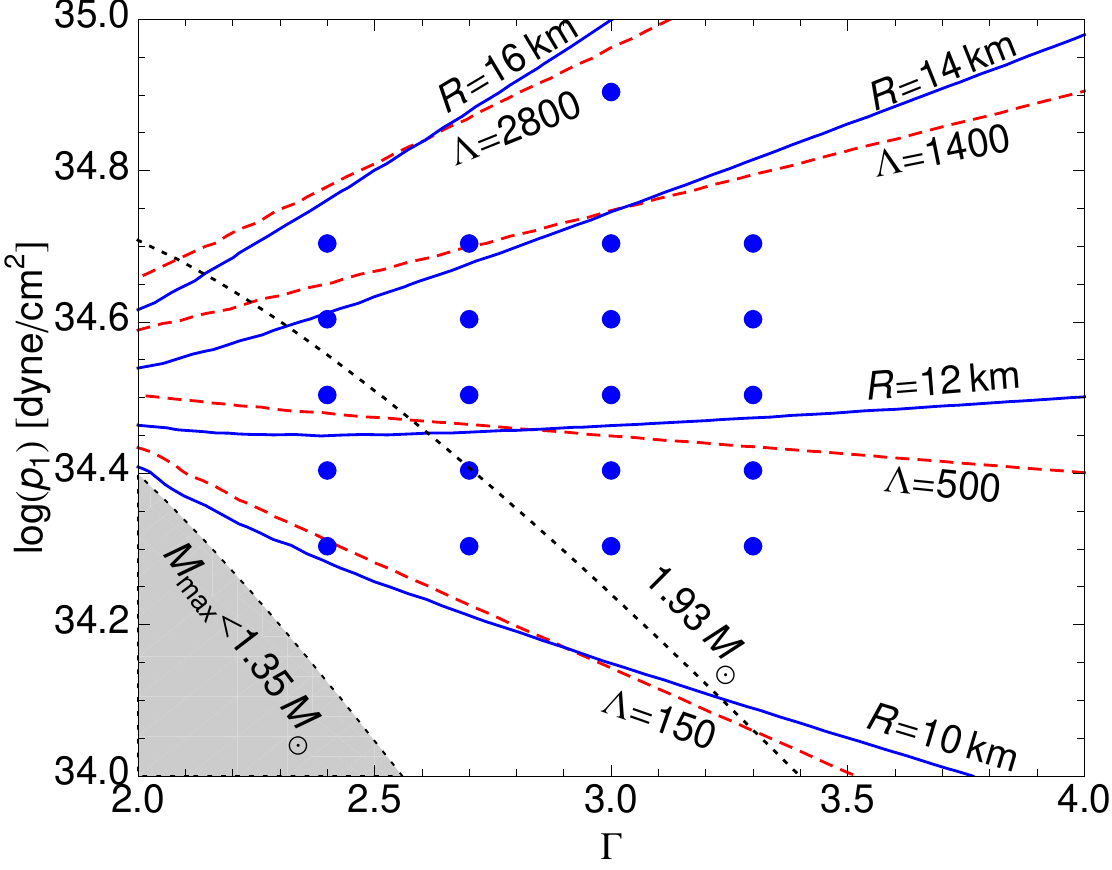}
\end{center}
\caption{ \label{fig:paramspace}
The 21 EOS used in the simulations are represented by blue points in the parameter space.  For a NS of mass 1.35~$M_\odot$, NS radius contours are solid blue and tidal deformability contours are dashed red.  Also shown are two dotted contours of maximum NS mass.  EOS parameters in the shaded region do not allow a 1.35~$M_\odot$ NS.}
\end{figure}

We have performed 134 simulations of the late inspiral, merger, and ringdown of BHNS systems, using 21 sets of parameters for our two-parameter EOS. We have also varied the mass ratio from $Q=2$ to 5, the spin of the black hole from $\chi_{\rm BH} = - 0.5$ to 0.75, and the neutron star mass from $M_{\rm NS} = 1.20$~$M_\odot$ to 1.45~$M_\odot$. The EOS parameters used as well as the corresponding NS radius, Love number, and tidal deformability for the three NS masses can be found in Table~\ref{tab:properties}. A list of all the simulations and their starting frequencies is given in Table~\ref{tab:simulations}. In addition, we plot the EOS as points in parameter space in Fig.~\ref{fig:paramspace} along with contours of radius, tidal deformability $\Lambda$, and maximum NS mass.  The 1.93~$M_\odot$ maximum mass contour corresponds to the constraint from the recently observed pulsar with a mass of $1.97\pm0.04~M_\odot$ measured using the Shapiro delay~\cite{DemorestPennucciRansom2010}.  For this parametrized EOS, parameters below this curve have a maximum mass less than 1.93~$M_\odot$, and therefore do not agree with the measurement of a NS with $M_{\rm NS}>1.93~M_\odot$.

\begin{table*}[!htb]
\caption[Neutron star properties for the 21 EOS used in the simulations]{ \label{tab:properties}
Neutron star properties for the 21 EOS used in the simulations.  The original EOS names~\cite{ReadMarkakisShibata2009, KyutokuShibataTaniguchi2010, KyutokuOkawaShibataTaniguchi2011} are also listed.  $p_1$ is given in units of~dyne/cm$^2$, maximum mass is in $M_\odot$, and neutron star radius $R$ is in~km.  $R$, $k_2$, and $\Lambda$ are given for the three masses used: $\{1.20, 1.35, 1.45\}~M_\odot$.
}
\begin{center}
\begin{tabular}{llccc|ccr|ccr|rcr}
\hline\hline
\multicolumn{2}{c}{EOS} & $\log p_1$ & $\Gamma$ & $M_{\rm max}$ & $R_{1.20}$ & $k_{2, 1.20}$ & $\Lambda_{1.20}$ & $R_{1.35}$ & $k_{2, 1.35}$ & $\Lambda_{1.35}$ & $R_{1.45}$ & $k_{2, 1.45}$ & $\Lambda_{1.45}$\\
\hline

$p.3\Gamma 2.4$ & Bss & 34.3 & 2.4 & 1.566 & 10.66 & 0.0765 & 401 & 10.27 & 0.0585 & 142 & 9.89 & 0.0455 & 64 \\
$p.3\Gamma 2.7$ & Bs & 34.3 & 2.7 & 1.799 & 10.88 & 0.0910 & 528 & 10.74 & 0.0751 & 228 & 10.61 & 0.0645 & 129 \\
$p.3\Gamma 3.0$ & B & 34.3 & 3.0 & 2.002 & 10.98 & 0.1010 & 614 & 10.96 & 0.0861 & 288 & 10.92 & 0.0762 & 176 \\
$p.3\Gamma 3.3$ & & 34.3 & 3.3 & 2.181 & 11.04 & 0.1083 & 677 & 11.09 & 0.0941 & 334 & 11.10 & 0.0847 & 212 \\

$p.4\Gamma 2.4$ & HBss & 34.4 & 2.4 & 1.701 & 11.74 & 0.0886 & 755 & 11.45 & 0.0723 & 301 & 11.19 & 0.0610 & 158 \\
$p.4\Gamma 2.7$ & HBs & 34.4 & 2.7 & 1.925 & 11.67 & 0.1004 & 828 & 11.57 & 0.0855 & 375 & 11.47 & 0.0754 & 222 \\
$p.4\Gamma 3.0$ & HB & 34.4 & 3.0 & 2.122 & 11.60 & 0.1088 & 872 & 11.61 & 0.0946 & 422 & 11.59 & 0.0851 & 263 \\
$p.4\Gamma 3.3$ & & 34.4 & 3.3 & 2.294 & 11.55 & 0.1151 & 903 & 11.62 & 0.1013 & 454 & 11.65 & 0.0921 & 293 \\

$p.5\Gamma 2.4$ &  & 34.5 & 2.4 & 1.848 & 12.88 & 0.1000 & 1353 & 12.64 & 0.0850 & 582 & 12.44 & 0.0747 & 330 \\
$p.5\Gamma 2.7$ & & 34.5 & 2.7 & 2.061 & 12.49 & 0.1096 & 1271 & 12.42 & 0.0954 & 598 & 12.35 & 0.0859 & 366 \\
$p.5\Gamma 3.0$ & H & 34.5 & 3.0 & 2.249 & 12.25 & 0.1165 & 1225 & 12.27 & 0.1029 & 607 & 12.27 & 0.0937 & 387 \\
$p.5\Gamma 3.3$ & & 34.5 & 3.3 & 2.413 & 12.08 & 0.1217 & 1196 & 12.17 & 0.1085 & 613 & 12.21 & 0.0995 & 400 \\

$p.6\Gamma 2.4$ & & 34.6 & 2.4 & 2.007 & 14.08 & 0.1108 & 2340 & 13.89 & 0.0970 & 1061 & 13.73 & 0.0875 & 633 \\
$p.6\Gamma 2.7$ & & 34.6 & 2.7 & 2.207 & 13.35 & 0.1184 & 1920 & 13.32 & 0.1051 & 932 & 13.27 & 0.0960 & 585 \\
$p.6\Gamma 3.0$ & & 34.6 & 3.0 & 2.383 & 12.92 & 0.1240 & 1704 & 12.97 & 0.1110 & 862 & 12.98 & 0.1022 & 558 \\
$p.6\Gamma 3.3$ & & 34.6 & 3.3 & 2.537 & 12.63 & 0.1282 & 1575 & 12.74 & 0.1155 & 819 & 12.79 & 0.1068 & 541 \\

$p.7\Gamma 2.4$ & & 34.7 & 2.4 & 2.180 & 15.35 & 0.1210 & 3941 & 15.20 & 0.1083 & 1860 & 15.07 & 0.0995 & 1147 \\
$p.7\Gamma 2.7$ & & 34.7 & 2.7 & 2.362 & 14.26 & 0.1269 & 2859 & 14.25 & 0.1144 & 1423 & 14.22 & 0.1058 & 912 \\
$p.7\Gamma 3.0$ & 1.5H & 34.7 & 3.0 & 2.525 & 13.62 & 0.1313 & 2351 & 13.69 & 0.1189 & 1211 & 13.72 & 0.1104 & 795 \\
$p.7\Gamma 3.3$ & & 34.7 & 3.3 & 2.669 & 13.20 & 0.1346 & 2062 & 13.32 & 0.1223 & 1087 & 13.39 & 0.1140 & 726 \\

$p.9\Gamma 3.0$ & 2H & 34.9 & 3.0 & 2.834 & 15.12 & 0.1453 & 4382 & 15.22 & 0.1342 & 2324 & 15.28 & 0.1264 & 1560 \\

\hline\hline
\end{tabular}
\end{center}
\end{table*}

\begin{table*}[!htb]
\caption{ \label{tab:simulations}
Data for the 134 BHNS simulations.  NS mass is in units of $M_\odot$. $M\Omega_0$ is the initial orbital angular velocity of the system. The columns labeled ``Phen'' and ``EOB'' indicate if analytic PhenomC and EOB BBH waveforms are available for the given values of $\chi_{\rm BH}$ and $Q$. ``C'' indicates the waveform is available and is calibrated for those parameter values. ``NC'' indicates the waveform could be generated but is not calibrated. ``NA'' indicates the waveform is not available.
}
\begin{center}
\begin{tabular}{cccccrr|cccccrr|cccccrr}
\hline\hline
$\chi_{\rm BH}$ & $Q$ & $M_{\rm NS}$ & EOS & $M\Omega_0$ & Phen & EOB & $\chi_{\rm BH}$ & $Q$ & $M_{\rm NS}$ & EOS & $M\Omega_0$ & Phen & EOB & $\chi_{\rm BH}$ & $Q$ & $M_{\rm NS}$ & EOS & $M\Omega_0$ & Phen & EOB \\
\hline

-0.5 & 2 & 1.35 & $p.3 \Gamma 3.0$ & 0.028 & C & C & 0.25 & 3 & 1.35 & $p.7 \Gamma 3.0$ & 0.030 & C & C & 0.75 & 2 & 1.20 & $p.3 \Gamma 3.0$ & 0.028 & C & NA \\
-0.5 & 2 & 1.35 & $p.4 \Gamma 3.0$ & 0.028 & C & C & 0.25 & 3 & 1.35 & $p.9 \Gamma 3.0$ & 0.028 & C & C & 0.75 & 2 & 1.20 & $p.4 \Gamma 3.0$ & 0.028 & C & NA \\
-0.5 & 2 & 1.35 & $p.5 \Gamma 3.0$ & 0.025 & C & C & 0.25 & 4 & 1.35 & $p.3 \Gamma 3.0$ & 0.031 & C & C & 0.75 & 2 & 1.20 & $p.5 \Gamma 3.0$ & 0.028 & C & NA \\
-0.5 & 2 & 1.35 & $p.9 \Gamma 3.0$ & 0.022 & C & C & 0.25 & 4 & 1.35 & $p.5 \Gamma 3.0$ & 0.031 & C & C & 0.75 & 2 & 1.20 & $p.9 \Gamma 3.0$ & 0.025 & C & NA \\
-0.5 & 3 & 1.35 & $p.4 \Gamma 3.0$ & 0.030 & C & C & 0.25 & 4 & 1.35 & $p.7 \Gamma 3.0$ & 0.031 & C & C & 0.75 & 2 & 1.35 & $p.3 \Gamma 3.0$ & 0.028 & C & NA \\
0 & 2 & 1.20 & $p.3 \Gamma 3.0$ & 0.028 & C & C & 0.25 & 4 & 1.35 & $p.9 \Gamma 3.0$ & 0.029 & C & C & 0.75 & 2 & 1.35 & $p.4 \Gamma 3.0$ & 0.028 & C & NA \\
0 & 2 & 1.20 & $p.4 \Gamma 3.0$ & 0.028 & C & C & 0.25 & 5 & 1.35 & $p.3 \Gamma 3.0$ & 0.033 & NC & C & 0.75 & 2 & 1.35 & $p.5 \Gamma 3.0$ & 0.028 & C & NA \\
0 & 2 & 1.20 & $p.5 \Gamma 3.0$ & 0.028 & C & C & 0.25 & 5 & 1.35 & $p.5 \Gamma 3.0$ & 0.033 & NC & C & 0.75 & 2 & 1.35 & $p.7 \Gamma 3.0$ & 0.028 & C & NA \\
0 & 2 & 1.20 & $p.9 \Gamma 3.0$ & 0.022 & C & C & 0.25 & 5 & 1.35 & $p.7 \Gamma 3.0$ & 0.033 & NC & C & 0.75 & 2 & 1.35 & $p.9 \Gamma 3.0$ & 0.025 & C & NA \\
0 & 2 & 1.35 & $p.3 \Gamma 2.4$ & 0.028 & C & C & 0.25 & 5 & 1.35 & $p.9 \Gamma 3.0$ & 0.033 & NC & C & 0.75 & 2 & 1.45 & $p.3 \Gamma 3.0$ & 0.028 & C & NA \\
0 & 2 & 1.35 & $p.3 \Gamma 2.7$ & 0.028 & C & C & 0.5 & 2 & 1.35 & $p.3 \Gamma 3.0$ & 0.028 & C & C & 0.75 & 2 & 1.45 & $p.4 \Gamma 3.0$ & 0.028 & C & NA \\
0 & 2 & 1.35 & $p.3 \Gamma 3.0$ & 0.028 & C & C & 0.5 & 2 & 1.35 & $p.4 \Gamma 3.0$ & 0.028 & C & C & 0.75 & 2 & 1.45 & $p.5 \Gamma 3.0$ & 0.028 & C & NA \\
0 & 2 & 1.35 & $p.3 \Gamma 3.3$ & 0.025 & C & C & 0.5 & 2 & 1.35 & $p.5 \Gamma 3.0$ & 0.028 & C & C & 0.75 & 2 & 1.45 & $p.9 \Gamma 3.0$ & 0.025 & C & NA \\
0 & 2 & 1.35 & $p.4 \Gamma 2.4$ & 0.028 & C & C & 0.5 & 2 & 1.35 & $p.7 \Gamma 3.0$ & 0.028 & C & C & 0.75 & 3 & 1.35 & $p.3 \Gamma 3.0$ & 0.030 & C & NA \\
0 & 2 & 1.35 & $p.4 \Gamma 2.7$ & 0.028 & C & C & 0.5 & 2 & 1.35 & $p.9 \Gamma 3.0$ & 0.025 & C & C & 0.75 & 3 & 1.35 & $p.4 \Gamma 3.0$ & 0.030 & C & NA \\
0 & 2 & 1.35 & $p.4 \Gamma 3.0$ & 0.028 & C & C & 0.5 & 3 & 1.35 & $p.3 \Gamma 2.4$ & 0.030 & C & C & 0.75 & 3 & 1.35 & $p.5 \Gamma 3.0$ & 0.030 & C & NA \\
0 & 2 & 1.35 & $p.4 \Gamma 3.3$ & 0.025 & C & C & 0.5 & 3 & 1.35 & $p.3 \Gamma 2.7$ & 0.030 & C & C & 0.75 & 3 & 1.35 & $p.7 \Gamma 3.0$ & 0.030 & C & NA \\
0 & 2 & 1.35 & $p.5 \Gamma 2.4$ & 0.025 & C & C & 0.5 & 3 & 1.35 & $p.3 \Gamma 3.0$ & 0.030 & C & C & 0.75 & 3 & 1.35 & $p.9 \Gamma 3.0$ & 0.028 & C & NA \\
0 & 2 & 1.35 & $p.5 \Gamma 2.7$ & 0.025 & C & C & 0.5 & 3 & 1.35 & $p.3 \Gamma 3.3$ & 0.030 & C & C & 0.75 & 4 & 1.35 & $p.3 \Gamma 3.0$ & 0.032 & C & NA \\
0 & 2 & 1.35 & $p.5 \Gamma 3.0$ & 0.028 & C & C & 0.5 & 3 & 1.35 & $p.4 \Gamma 2.4$ & 0.030 & C & C & 0.75 & 4 & 1.35 & $p.4 \Gamma 3.0$ & 0.032 & C & NA \\
0 & 2 & 1.35 & $p.5 \Gamma 3.3$ & 0.025 & C & C & 0.5 & 3 & 1.35 & $p.4 \Gamma 2.7$ & 0.030 & C & C & 0.75 & 4 & 1.35 & $p.5 \Gamma 3.0$ & 0.032 & C & NA \\
0 & 2 & 1.35 & $p.6 \Gamma 2.4$ & 0.025 & C & C & 0.5 & 3 & 1.35 & $p.4 \Gamma 3.0$ & 0.030 & C & C & 0.75 & 4 & 1.35 & $p.7 \Gamma 3.0$ & 0.032 & C & NA \\
0 & 2 & 1.35 & $p.6 \Gamma 2.7$ & 0.025 & C & C & 0.5 & 3 & 1.35 & $p.4 \Gamma 3.3$ & 0.030 & C & C & 0.75 & 4 & 1.35 & $p.9 \Gamma 3.0$ & 0.030 & C & NA \\
0 & 2 & 1.35 & $p.6 \Gamma 3.0$ & 0.025 & C & C & 0.5 & 3 & 1.35 & $p.5 \Gamma 2.4$ & 0.030 & C & C & 0.75 & 5 & 1.35 & $p.3 \Gamma 2.4$ & 0.036 & NC & NA \\
0 & 2 & 1.35 & $p.6 \Gamma 3.3$ & 0.025 & C & C & 0.5 & 3 & 1.35 & $p.5 \Gamma 2.7$ & 0.030 & C & C & 0.75 & 5 & 1.35 & $p.3 \Gamma 2.7$ & 0.036 & NC & NA \\
0 & 2 & 1.35 & $p.7 \Gamma 2.4$ & 0.025 & C & C & 0.5 & 3 & 1.35 & $p.5 \Gamma 3.0$ & 0.030 & C & C & 0.75 & 5 & 1.35 & $p.3 \Gamma 3.0$ & 0.036 & NC & NA \\
0 & 2 & 1.35 & $p.7 \Gamma 2.7$ & 0.025 & C & C & 0.5 & 3 & 1.35 & $p.5 \Gamma 3.3$ & 0.030 & C & C & 0.75 & 5 & 1.35 & $p.3 \Gamma 3.3$ & 0.036 & NC & NA \\
0 & 2 & 1.35 & $p.7 \Gamma 3.0$ & 0.028 & C & C & 0.5 & 3 & 1.35 & $p.6 \Gamma 2.4$ & 0.030 & C & C & 0.75 & 5 & 1.35 & $p.4 \Gamma 2.4$ & 0.036 & NC & NA \\
0 & 2 & 1.35 & $p.7 \Gamma 3.3$ & 0.025 & C & C & 0.5 & 3 & 1.35 & $p.6 \Gamma 2.7$ & 0.030 & C & C & 0.75 & 5 & 1.35 & $p.4 \Gamma 2.7$ & 0.036 & NC & NA \\
0 & 2 & 1.35 & $p.9 \Gamma 3.0$ & 0.025 & C & C & 0.5 & 3 & 1.35 & $p.6 \Gamma 3.0$ & 0.030 & C & C & 0.75 & 5 & 1.35 & $p.4 \Gamma 3.0$ & 0.036 & NC & NA \\
0 & 3 & 1.35 & $p.3 \Gamma 3.0$ & 0.030 & C & C & 0.5 & 3 & 1.35 & $p.6 \Gamma 3.3$ & 0.030 & C & C & 0.75 & 5 & 1.35 & $p.4 \Gamma 3.3$ & 0.036 & NC & NA \\
0 & 3 & 1.35 & $p.4 \Gamma 3.0$ & 0.030 & C & C & 0.5 & 3 & 1.35 & $p.7 \Gamma 2.4$ & 0.028 & C & C & 0.75 & 5 & 1.35 & $p.5 \Gamma 2.4$ & 0.036 & NC & NA \\
0 & 3 & 1.35 & $p.5 \Gamma 3.0$ & 0.030 & C & C & 0.5 & 3 & 1.35 & $p.7 \Gamma 2.7$ & 0.028 & C & C & 0.75 & 5 & 1.35 & $p.5 \Gamma 2.7$ & 0.036 & NC & NA \\
0 & 3 & 1.35 & $p.7 \Gamma 3.0$ & 0.030 & C & C & 0.5 & 3 & 1.35 & $p.7 \Gamma 3.0$ & 0.030 & C & C & 0.75 & 5 & 1.35 & $p.5 \Gamma 3.0$ & 0.036 & NC & NA \\
0 & 3 & 1.35 & $p.9 \Gamma 3.0$ & 0.028 & C & C & 0.5 & 3 & 1.35 & $p.7 \Gamma 3.3$ & 0.030 & C & C & 0.75 & 5 & 1.35 & $p.5 \Gamma 3.3$ & 0.036 & NC & NA \\
0 & 4 & 1.35 & $p.3 \Gamma 3.0$ & 0.031 & C & C & 0.5 & 3 & 1.35 & $p.9 \Gamma 3.0$ & 0.028 & C & C & 0.75 & 5 & 1.35 & $p.6 \Gamma 2.4$ & 0.036 & NC & NA \\
0 & 4 & 1.35 & $p.5 \Gamma 3.0$ & 0.031 & C & C & 0.5 & 4 & 1.35 & $p.3 \Gamma 3.0$ & 0.025 & C & C & 0.75 & 5 & 1.35 & $p.6 \Gamma 2.7$ & 0.036 & NC & NA \\
0 & 4 & 1.35 & $p.7 \Gamma 3.0$ & 0.031 & C & C & 0.5 & 4 & 1.35 & $p.4 \Gamma 3.0$ & 0.035 & C & C & 0.75 & 5 & 1.35 & $p.6 \Gamma 3.0$ & 0.036 & NC & NA \\
0 & 4 & 1.35 & $p.9 \Gamma 3.0$ & 0.029 & C & C & 0.5 & 4 & 1.35 & $p.5 \Gamma 3.0$ & 0.035 & C & C & 0.75 & 5 & 1.35 & $p.6 \Gamma 3.3$ & 0.036 & NC & NA \\
0 & 5 & 1.35 & $p.3 \Gamma 3.0$ & 0.033 & NC & C & 0.5 & 4 & 1.35 & $p.7 \Gamma 3.0$ & 0.035 & C & C & 0.75 & 5 & 1.35 & $p.7 \Gamma 2.4$ & 0.036 & NC & NA \\
0 & 5 & 1.35 & $p.5 \Gamma 3.0$ & 0.033 & NC & C & 0.5 & 4 & 1.35 & $p.9 \Gamma 3.0$ & 0.035 & C & C & 0.75 & 5 & 1.35 & $p.7 \Gamma 2.7$ & 0.036 & NC & NA \\
0 & 5 & 1.35 & $p.7 \Gamma 3.0$ & 0.033 & NC & C & 0.5 & 5 & 1.35 & $p.3 \Gamma 3.0$ & 0.033 & NC & C & 0.75 & 5 & 1.35 & $p.7 \Gamma 3.0$ & 0.036 & NC & NA \\
0 & 5 & 1.35 & $p.9 \Gamma 3.0$ & 0.031 & NC & C & 0.5 & 5 & 1.35 & $p.5 \Gamma 3.0$ & 0.033 & NC & C & 0.75 & 5 & 1.35 & $p.7 \Gamma 3.3$ & 0.036 & NC & NA \\
0.25 & 3 & 1.35 & $p.3 \Gamma 3.0$ & 0.030 & C & C & 0.5 & 5 & 1.35 & $p.7 \Gamma 3.0$ & 0.033 & NC & C & 0.75 & 5 & 1.35 & $p.9 \Gamma 3.0$ & 0.036 & NC & NA \\
0.25 & 3 & 1.35 & $p.5 \Gamma 3.0$ & 0.030 & C & C & 0.5 & 5 & 1.35 & $p.9 \Gamma 3.0$ & 0.033 & NC & C & &&&&&&\\

\hline\hline
\end{tabular}
\end{center}
\end{table*}

\begin{figure}[!htb]
\begin{center}
\includegraphics[width=85mm]{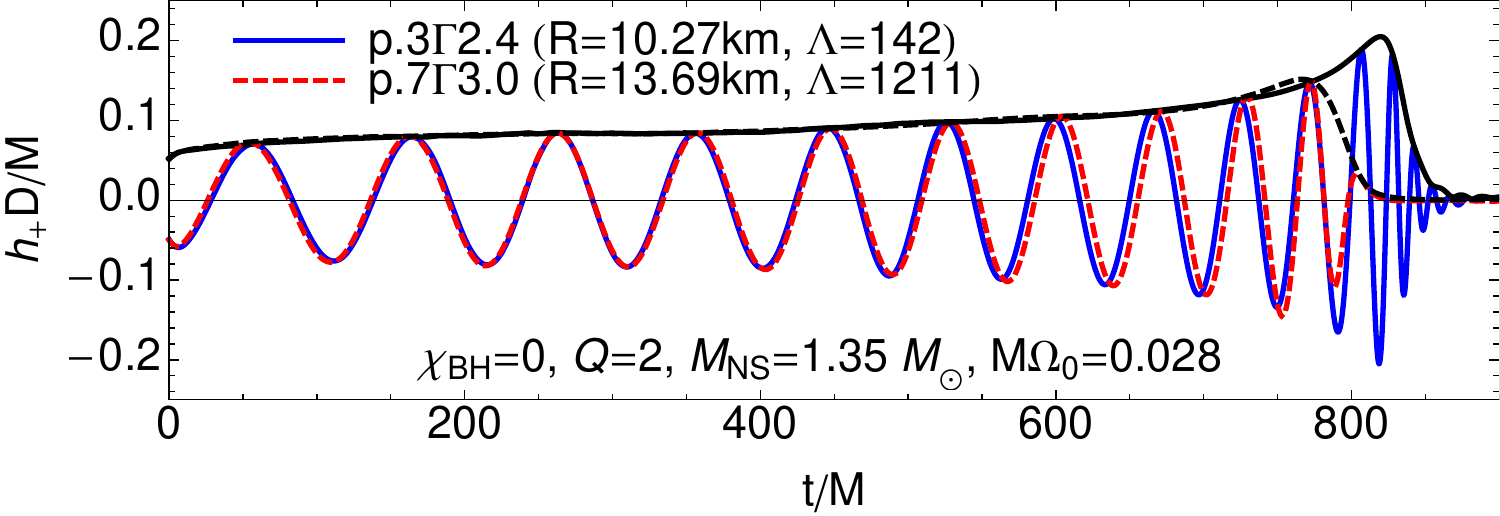}
\end{center}
\caption{ \label{fig:hoft}
Polarization $h_+$ evaluated on the orbital axis for two waveforms from numerical BHNS simulations that differ only in their EOS. Also shown in black are the waveform amplitudes $|h_+ - ih_\times|$.}
\end{figure}

In Fig.~\ref{fig:hoft} we show two representative waveforms. The waveform with a very soft EOS ($p.3\Gamma2.4$), and therefore small radius and tidal deformability, behaves very much like a BBH waveform where the inspiral smoothly transitions to quasinormal mode ringdown. For the stiff EOS ($p.7\Gamma3.0$), however, the neutron star is tidally disrupted near the end of inspiral; the disruption and the spread of tidally stripped matter to form a roughly axisymmetric disk leads to a rapid decrease in the waveform amplitude and suppresses the subsequent ringdown.

\begin{figure*}[!htb]
\begin{center}
\includegraphics[width=2.9in]{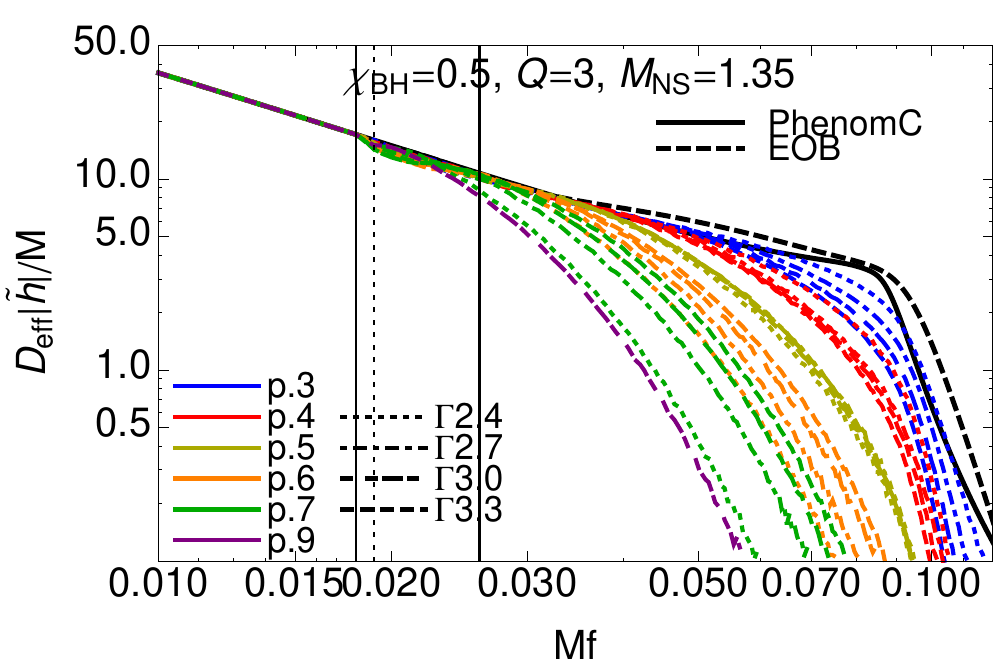}
\includegraphics[width=2.9in]{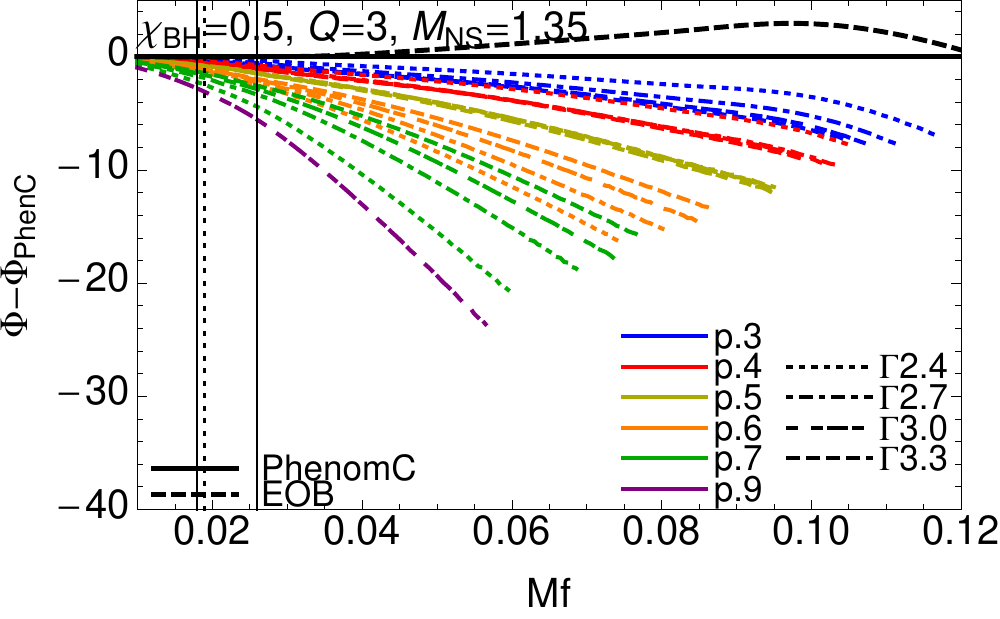}
\end{center}
\caption{ \label{fig:hybridvarylambda}
(Color online) Left panel: Amplitude $|\tilde h|$ of the Fourier transform of BHNS waveforms for $\chi_{\rm BH} = 0.5$, $Q = 3$, and $M_{\rm NS} = 1.35M_\odot$ and 21 EOS. Color indicates the value of $\log(p_1)$ while the line style indicates the value of $\Gamma$. Also shown are two analytic approximations to the BBH waveform with the same values of $\chi_{\rm BH}$ and $Q$: PhenomC and EOB discussed in Section~\ref{sec:hybrid}. Right panel: Phase of the Fourier transformed waveform relative to the PhenomC BBH waveform $\Delta\Phi = \Phi - \Phi_{\rm Phen}$. The curves are truncated when the amplitude $D_{\rm eff} |\tilde h| / M$ drops below 0.1, and numerical error begins to dominate. The phase of the EOB BBH waveform is also shown relative to the PhenomC BBH waveform. In both figures the BHNS and EOB waveforms have been windowed, matched, and spliced to the tidally corrected PhenomC waveform as described in Section~\ref{sec:hybrid}. The windowing width $\Delta t_{\rm win} = 300M$; the start and end frequencies for matching, represented by solid vertical lines, are $Mf_i = 0.018$ and $Mf_f = 0.026$; and the splicing interval, represented by dotted vertical lines, is $Ms_i = Mf_i$ (overlapping with the solid line) and $Ms_f = Ms_i + 0.001$.}
\end{figure*}

\begin{figure*}[!htb]
\begin{center}
\includegraphics[width=2.9in]{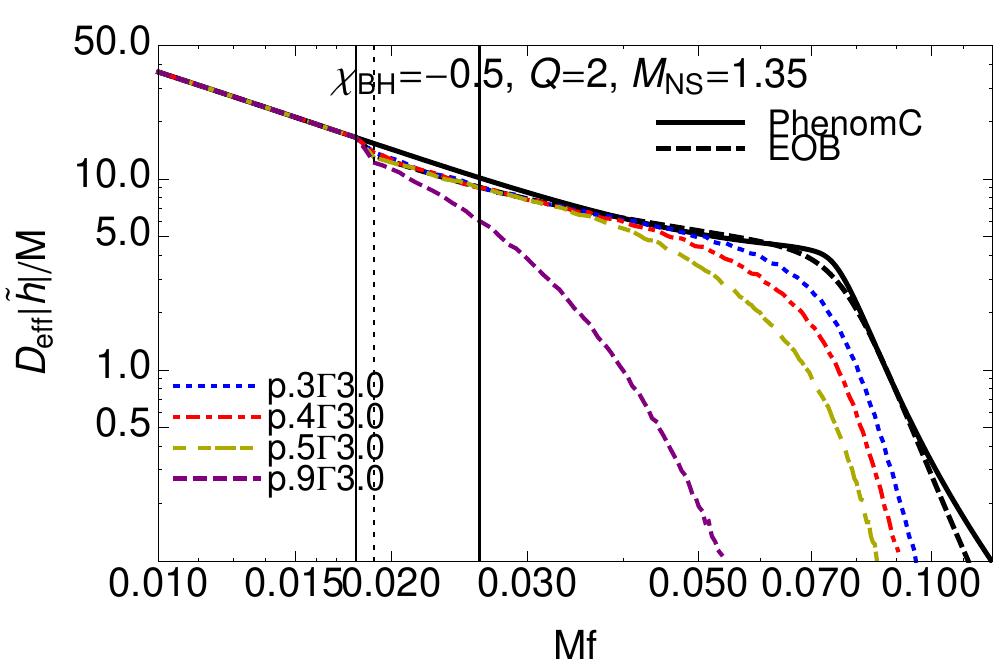}
\includegraphics[width=2.9in]{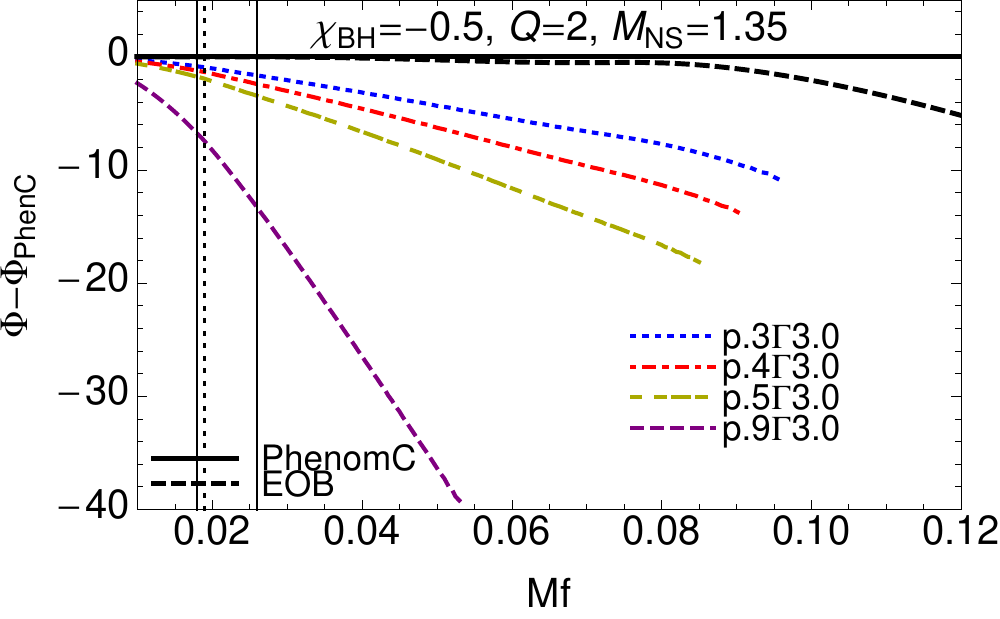}
\end{center}
\caption{ \label{fig:hybridvarylambda2}
Same as Fig.~\ref{fig:hybridvarylambda}, but $\chi_{\rm BH} = -0.5$, $Q = 2$, and $M_{\rm NS} = 1.35M_\odot$.}
\end{figure*}

\begin{figure*}[!htb]
\begin{center}
\includegraphics[width=2.9in]{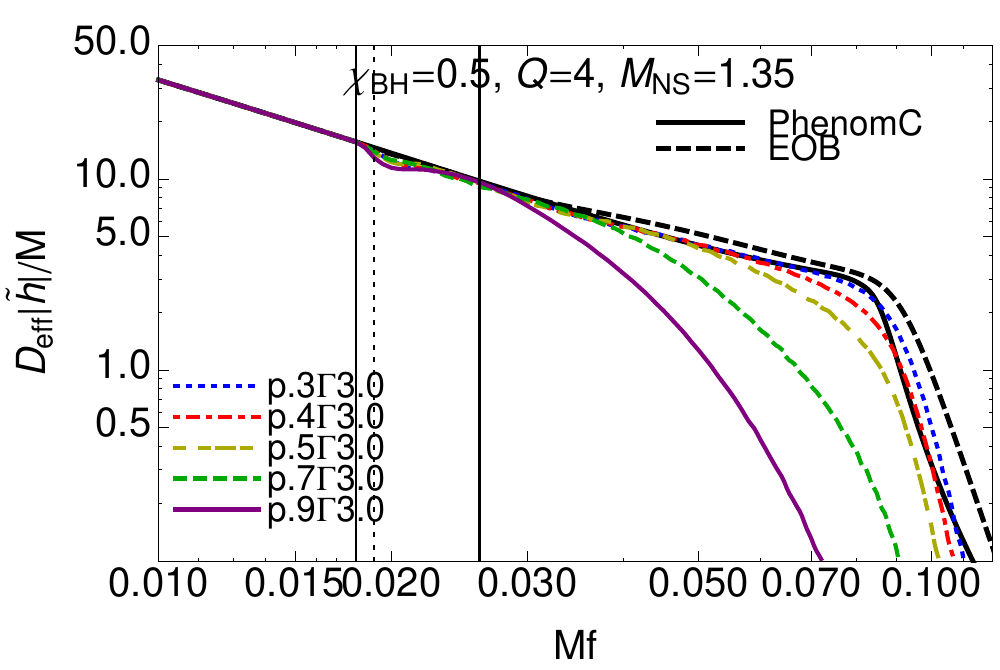}
\includegraphics[width=2.9in]{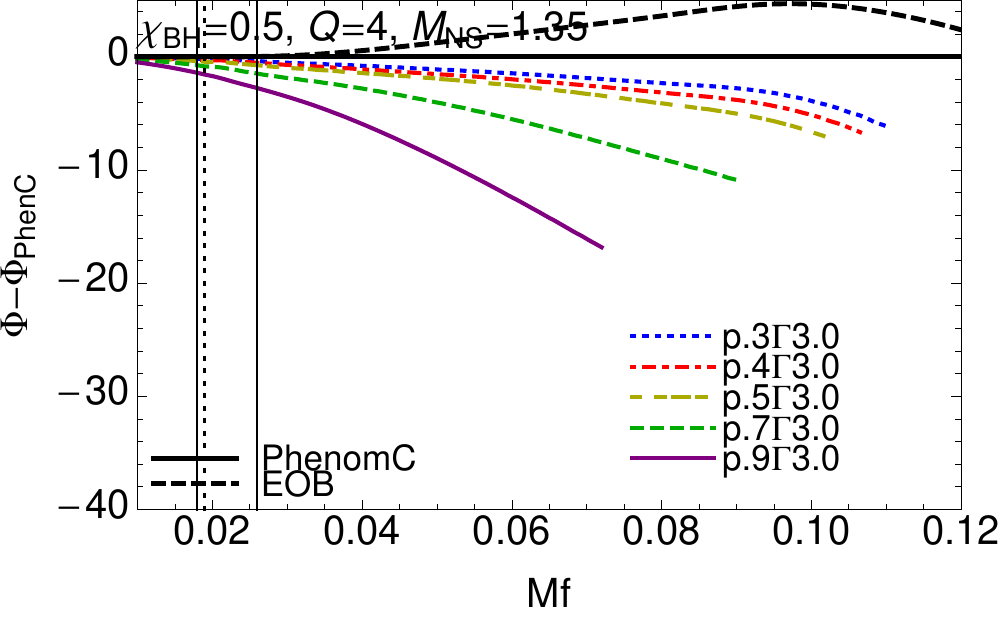}
\end{center}
\caption{ \label{fig:hybridvarylambda3}
Same as Fig.~\ref{fig:hybridvarylambda}, but $\chi_{\rm BH} = 0.5$, $Q = 4$, and $M_{\rm NS} = 1.35M_\odot$. During the ringdown, the PhenomC waveform amplitude is noticeably less than the BHNS waveform with the softest EOS ($p.3\Gamma3.0$) around $Mf=0.1$. The EOB amplitude, however, is always greater than the BHNS amplitude.}
\end{figure*}

\begin{figure*}[!htb]
\begin{center}
\includegraphics[width=2.9in]{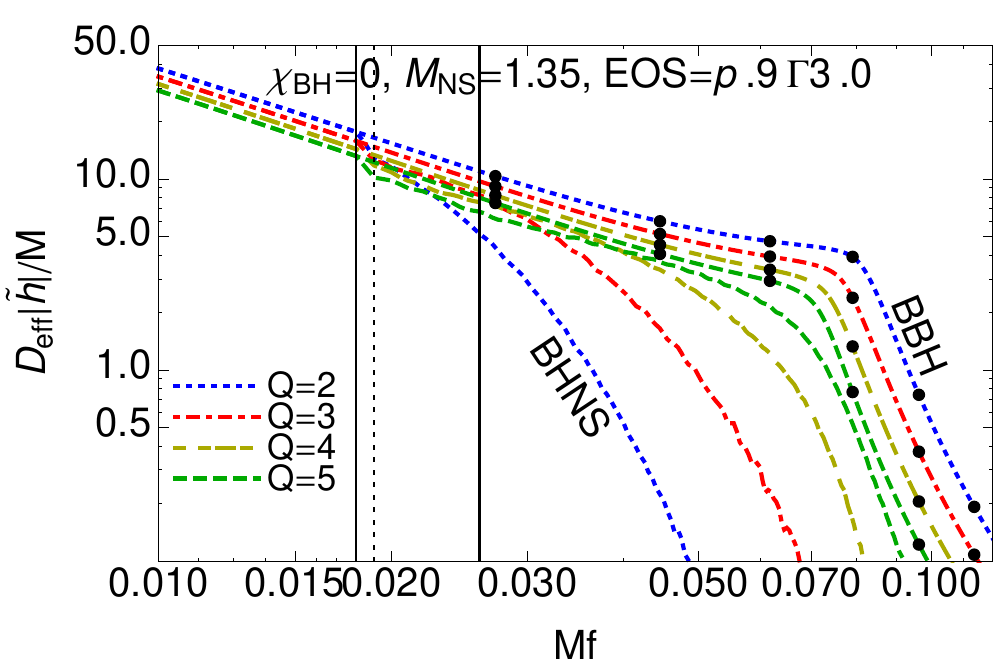}
\includegraphics[width=2.9in]{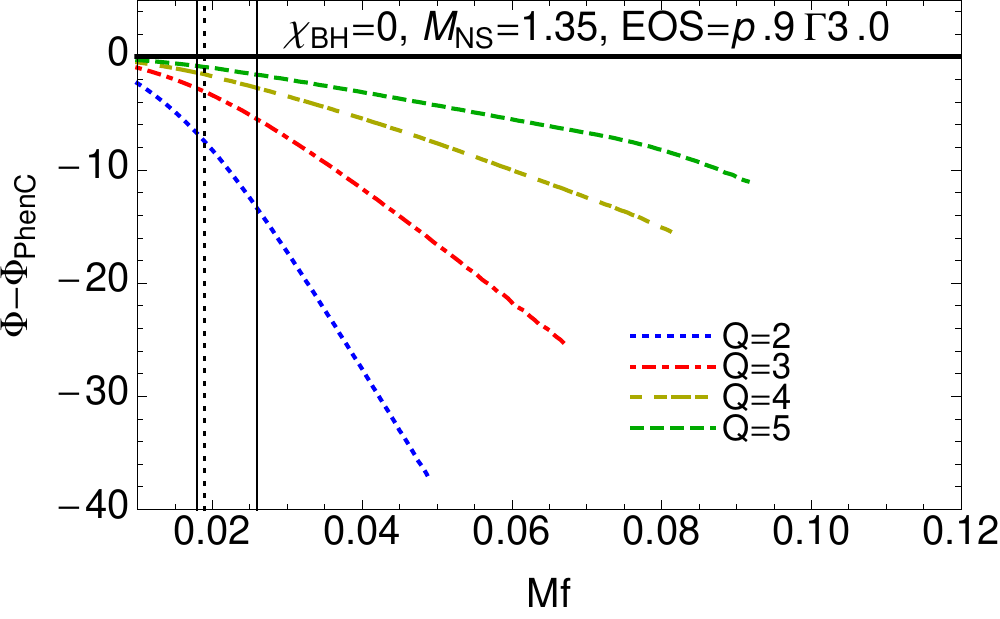}
\end{center}
\caption{ \label{fig:hybridvaryq}
BHNS waveforms with mass ratios of $Q = \{2, 3, 4, 5\}$ with all other parameters fixed at $\chi_{\rm BH} = 0$, $M_{\rm NS} = 1.35M_\odot$, and the stiffest $\mathrm{EOS} = p.9\Gamma3.0$. Left panel: Amplitude of BHNS waveform as well as PhenomC BBH waveforms (black dots) with the same values of $Q$ and $\chi_{\rm BH}$. The difference in amplitude between BBH and BHNS waveform decreases with the mass ratio $Q$. Right panel: Difference in phase between BHNS and PhenomC BBH waveform also decreases with $Q$. Windowing, matching, and splicing are identical to Fig.~\ref{fig:hybridvarylambda}.}
\end{figure*}

\begin{figure*}[!htb]
\begin{center}
\includegraphics[width=2.9in]{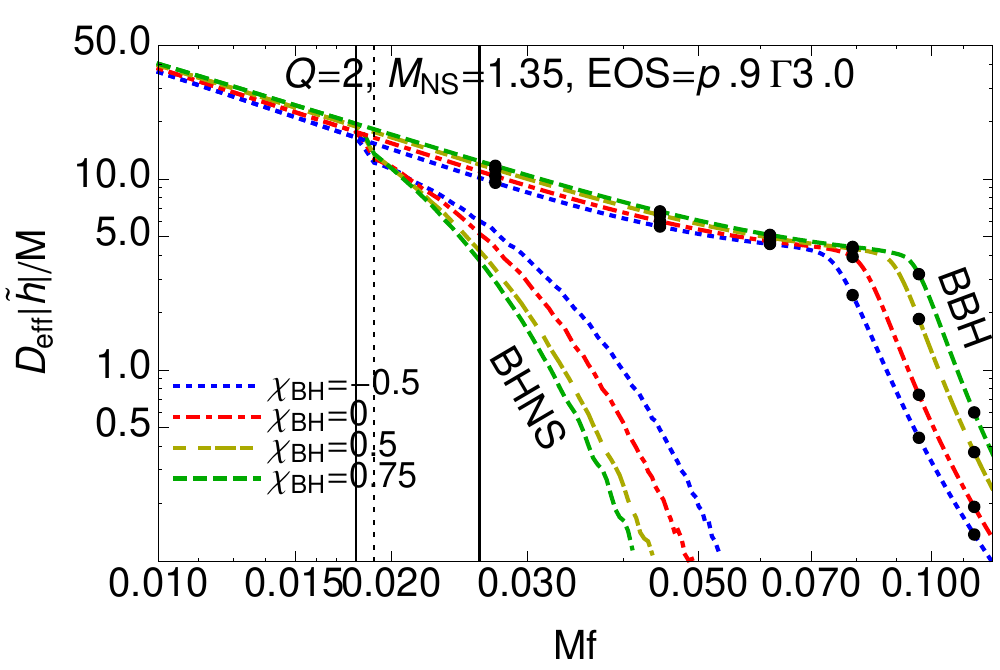}
\includegraphics[width=2.9in]{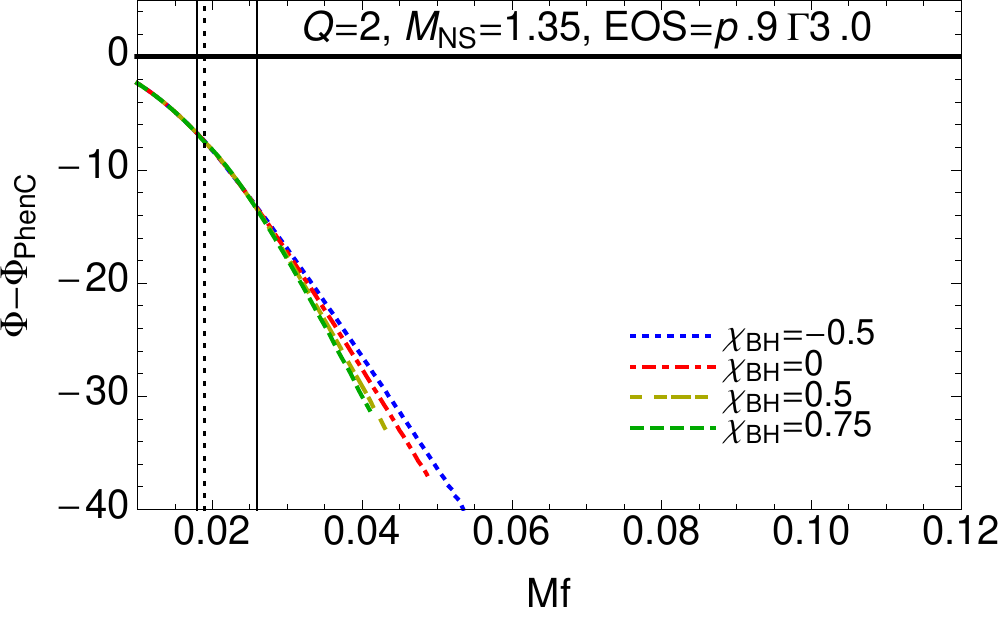}
\end{center}
\caption{ \label{fig:hybridvarychi}
Same as Fig.~\ref{fig:hybridvaryq} except BH spin is varied instead of mass ratio. Waveforms with spins of $\chi_{\rm BH} = \{-0.5, 0, 0.5, 0.75\}$ are shown, and $Q = 2$, $M_{\rm NS} = 1.35M_\odot$, $\mathrm{EOS} = p.9\Gamma3.0$ for all waveforms. In contrast with the mass ratio (Fig.~\ref{fig:hybridvaryq}), the departure from the BBH waveform increases as the spin increases. However, the effect is moderate.}
\end{figure*}

Because trends in the BHNS waveform are most apparent in terms of the amplitude and phase of the Fourier transform, and because data analysis is usually done in the frequency domain, we will now focus our discussion of the waveforms on the frequency-domain waveform behavior. Several representative waveforms with varying tidal deformability $\Lambda$, mass ratio $Q$, and spin $\chi_{\rm BH}$ are shown in Figs.~\ref{fig:hybridvarylambda}--\ref{fig:hybridvarychi}.

As was found in Paper~I, the waveform monotonically departs from a BBH ($\Lambda = 0$) waveform as $\Lambda$ increases, and this is true for systems with spinning black holes as well, as we see from Figs.~\ref{fig:hybridvarylambda}--\ref{fig:hybridvarylambda3}. In particular, the cutoff frequency, where the waveform begins a sharp drop in the amplitude, decreases monotonically with increasing $\Lambda$. The accumulated BHNS phase $\Phi_{\rm BHNS}$ at fixed frequency $f$ similarly decreases with increasing $\Lambda$, because the orbit loses energy more rapidly: There is less time for the phase to accumulate.  As a result, the departure of $\Phi_{\rm BHNS}$ from the accumulated BBH phase $\Phi_{\rm BBH}$ increases with increasing $\Lambda$.

More massive black holes exert smaller tidal forces on their companion near coalescence, because the radius of the innermost orbit is roughly proportional to $M_{\rm BH}$.  As a result, the difference in amplitude and phase between a BHNS and BBH waveform decrease when the mass ratio $Q$ increases. The effect is clear in Fig.~\ref{fig:hybridvaryq}, which displays the dramatically enhanced departure of amplitude and phase from that of a BBH waveform as $Q$ decreases. On the other hand, the radius of the innermost orbit decreases with increasing aligned BH spin $\chi_{\rm BH}$, implying a larger maximum tidal force for larger $\chi_{\rm BH}$. The resulting enhanced departure from a BBH waveform is shown in Fig.~\ref{fig:hybridvarychi}.

\clearpage

\section{Constructing hybrid inspiral-merger-ringdown waveforms}
\label{sec:3}
\label{sec:hybrid}

To obtain as much information as possible about the physical parameters of a BHNS coalescence, we will construct a hybrid waveform that joins the numerical merger and ringdown waveforms to an analytic BHNS inspiral waveform. This inspiral waveform needs to account for the aligned spin $\chi_{\rm BH}$ of the BH as well as tidal interactions through the parameter $\Lambda$. In addition, because the numerical waveforms only include the last $\sim 10$ GW cycles before merger, the inspiral waveform, including spin and tidal corrections, should be valid as close to merger as possible. Finally, when we construct an analytic BHNS inspiral-merger-ringdown (IMR) waveform model in Section~\ref{sec:phenomBHNS}, we will find it useful to proceed by modifying a BBH IMR waveform, and so the waveform model will need to accurately model the merger and ringdown of a BBH system as well.

We will use two classes of waveform models that have been calibrated to numerical BBH simulations. The primary waveform model, described in the next subsection, is the frequency-domain BBH waveform labeled PhenomC~\cite{Santamaria2010}, and we will add a 1PN tidal correction to the inspiral portion of this waveform. We will also use a time-domain effective one body (EOB) waveform that incorporates spin corrections~\cite{TaracchiniPanBuonanno2012}, and we will again add tidal corrections to the inspiral. We compare these two waveform models with no inspiral tidal correction to each other in Figs.~\ref{fig:hybridvarylambda}--\ref{fig:hybridvarylambda3}. We note that although these two waveforms agree well for small mass ratios and spin, their differences become important as the mass ratio and spin increase.

\subsection{PhenomC waveform with tidal corrections}

Several frequency-domain phenomenological models are now available for the complete IMR BBH waveform. These models include the PhenomA~\cite{Ajith2008} model for nonspinning BBH systems, as well as the PhenomB~\cite{Ajith2011} model and improved PhenomC~\cite{Santamaria2010} model for aligned-spin BBH systems which we will use. 

In the PhenomC waveform, the Fourier transform of the waveform is decomposed into an amplitude $A_{\rm phen}(Mf)$ and phase $\Phi_{\rm phen}(Mf)$ as
\begin{equation}
\tilde h_{\rm phen}(Mf) = A_{\rm phen}(Mf) e^{ i \Phi_{\rm phen}(Mf) }.
\end{equation}
The inspiral is described by the stationary phase approximation TaylorF2 post-Newtonian waveform, and the spin contribution to the waveform of both bodies is approximated by the mass-weighted average spin~\cite{Santamaria2010}
\begin{equation}
\label{eq:chiavg}
\chi_{\rm avg} = \frac{M_1}{M} \chi_1 + \frac{M_2}{M} \chi_2.
\end{equation}
For our BHNS systems where the NS spin is assumed to be negligible compared to the BH spin, $\chi_{\rm avg} = (M_{\rm BH} / M) \chi_{\rm BH}$. The amplitude and phase of the merger and ringdown for frequencies above $Mf = 0.01$ are then fit to a large set of numerical simulations with mass ratios from 1--4 and various combinations of aligned or anti-aligned BH spins up to $|\chi_i| = 0.85$ as described in Ref.~\cite{Santamaria2010}.

We add tidal corrections to the inspiral for this waveform using the TaylorF2 stationary phase approximation up to 1PN order~\cite{VinesFlanaganHinderer2011}. This is the same quantity used by Pannarale et al.~\cite{PannaraleRezzollaOhmeRead2011} who found that BBH and BHNS waveforms are indistinguishable by aLIGO when only considering the inspiral. Explicitly, we add a tidal correction term $\psi_T(Mf)$ to the BBH phase,
\begin{equation}
\label{eq:1pntaylorf2}
\psi_T(Mf) = \frac{3}{128\eta} (\pi M f)^{-5/3} \left[ a_0 (\pi M f)^{10/3} + a_1 (\pi M f)^{12/3} \right],
\end{equation}
where in terms of the symmetric mass ratio $\eta = M_{\rm BH}M_{\rm NS}/(M_{\rm BH}+M_{\rm NS})^2$ and for $M_{\rm BH} \ge M_{\rm NS}$,
\begin{align}
\begin{split}
a_0 &= -12\left[ \left(1 + 7\eta - 31\eta^2 \right) \vphantom{\sqrt{1}} \right. \\
&- \left. \sqrt{1 - 4\eta} \left(1 + 9\eta  - 11\eta^2 \right) \right]\Lambda, \\
a_1 &= - \frac{585}{28} \left[ \left(1 + \frac{3775}{234}\eta - \frac{389}{6}\eta^2 + \frac{1376}{117}\eta^3 \right) \right. \\
&- \left. \sqrt{1 - 4\eta} \left(1 + \frac{4243}{234}\eta  - \frac{6217}{234}\eta^2 - \frac{10}{9}\eta^3 \right) \right]\Lambda. 
\end{split}
\end{align}
The inspiral waveform is now $\tilde h_{\rm insp}(Mf) = A_{\rm phen}(Mf) e^{i \Phi_{\rm phen+T}(Mf)}$, where $\Phi_{\rm phen+T}(Mf) = \Phi_{\rm phen}(Mf) + \psi_T(Mf)$.

We will work mostly with this frequency-domain waveform in the sections that follow. Since a BHNS waveform enters the detector band starting at frequencies as low as 10~Hz for aLIGO and 1~Hz for ET, it is much more efficient to start with a frequency-domain inspiral waveform than to evaluate the Fourier transform of a time-domain waveform.

\subsection{Spinning EOB waveform with tidal corrections}
\label{sec:spinEOB}

The other waveform model we will use, time-domain EOB waveforms, have proven succesful at reproducing the complete IMR of nonspinning BBH waveforms~\cite{DamourNagar2009}, and we used these EOB waveforms in Paper I to generate hybrid waveforms that did not incorporate tidal corrections for the inspiral. (See Appendix C of Paper I and references therein for a review of the EOB formalism.) Recently, spin terms have been calculated for the EOB Hamiltonian and resummed waveforms, and free parameters for the merger have been calibrated to numerical nonspinning BBH waveforms for mass ratios from 1--6, as well as for equal mass, aligned-spin waveforms with spins of $\chi_1 = \chi_2 = \pm 0.44$~\cite{TaracchiniPanBuonanno2012}. We will use EOB waveform tables generated by Taracchini and Buonanno~\cite{BuonannoTaracchini2012PC}. These tables were generated by evolving the EOB equations of motion with an initial radial coordinate of $r = 40M$ and a value of the radial velocity $\dot r$ consistent with the radiation reaction force to minimize initial eccentricity~\cite{BuonannoDamour2000}. The waveform is then evaluated starting at $r = 30M$ where any residual eccentricity is negligible~\cite{BuonannoTaracchini2012PC}.

In addition, tidal interactions have also recently been incorporated in the EOB formalism. In the method proposed in Ref.~\cite{DamourNagarVillain2012}, a term representing the conservative part of the tidal interaction is added to the radial potential $A(r)$ in the EOB Hamiltonian. Tidal corrections are also added to the resummed waveform $h_{\ell m}$ which are used to calculate the radiation reaction force in the equations of motion. The solutions to the equations of motion are then plugged back into $h_{\ell m}$ to produce a final waveform as a function of time. However, for simplicity and because the versions of the EOB formalism that incorporate spin and tidal interactions are slightly different and have not been calibrated to simulations with both spin and matter, we will instead incorporate tidal interactions in the spinning EOB waveform using the same method as that for the PhenomC waveform. Specifically, we Fourier transform the EOB waveform, decompose it into amplitude $A_{\rm EOB}(Mf)$ and phase $\Phi_{\rm EOB}(Mf)$, and then simply add the expression $\psi_T(Mf)$ from Eq.~\eqref{eq:1pntaylorf2} to $\Phi_{\rm EOB}(Mf)$. Future work on spinning BHNS systems should treat the spin and tidal interactions consistently.

\subsection{Hybridization in frequency domain}
\label{sec:hybridprocedure}

In Paper I, where we examined EOS effects only during the BHNS merger and ringdown, we used a time-domain matching method. In this paper, however, because we examine tidal effects for the entire IMR waveform, we will find it convenient to start with the inspiral waveform in the frequency domain before matching. The method we use closely follows the frequency-domain least-squares method used in Ref.~\cite{Santamaria2010} for BBH waveforms. 

We begin by windowing the numerical BHNS waveform with a Hann window over the interval $w_i$ to $w_f$ (width $\Delta t_{\rm win} = w_f - w_i$)
\begin{equation}
\label{eq:window}
w_{\rm on}(t) = \frac{1}{2}\left[1-\cos\left(\frac{\pi [t - w_i]}{w_f - w_i}\right)\right],
\end{equation}
and we choose the start of the windowing to be the start of the numerical waveform at $w_i = 0$. This windowing minimizes the oscillatory Gibbs phenomenon that results from Fourier transforming a waveform segment with nonzero starting amplitude. When matching waveforms, a time constant $\tau$ and phase constant $\phi$ are free parameters that need to be fixed. For a generic waveform $h(t)$, the time and phase can be adjusted to produce a shifted waveform $h^{\rm shift}(t; \tau, \phi) = h(t-\tau)e^{i \phi}$. The Fourier transformed waveform, which can be written in terms of amplitude and phase as $\tilde h(f) = |\tilde h(f)| e^{i\Phi(f)}$, has a corresponding shifted waveform $\tilde h^{\rm shift}(f; \tau, \phi) = |\tilde h(f)| e^{i\Phi^{\rm shift}(f; \tau, \phi)}$, where $\Phi^{\rm shift}(f; \tau, \phi) = \Phi(f) + 2\pi f \tau + \phi$. When joining the inspiral and numerical waveforms, we hold the phase of the tidally corrected inspiral waveform $\Phi_{\rm BBH+T}(f)$ fixed and adjust the phase of the numerical waveform $\Phi_{\rm NR}(f)$, such that $\Phi_{\rm NR}^{\rm shift}(f; \tau, \phi) = \Phi_{\rm NR}(f) + 2 \pi f \tau + \phi$. We then match the waveforms by performing a least-squares fit in the matching interval $f_i < f < f_f$ (width $\Delta f_{\rm match}$) that minimizes the quantity
\begin{equation}
\label{eq:leastsquare}
\int_{f_i}^{f_f} [ \Phi_{\rm NR}^{\rm shift}(f; \tau, \phi) - \Phi_{\rm BBH+T}(f) ]^2\, df
\end{equation}
to determine the free parameters $\tau$ and $\phi$. 

Once the time and phase shifts are found, we smoothly turn on the numerical waveform and smoothly turn off the phenomenological waveform within a splicing window $s_i < f < s_f$ (width $\Delta f_{\rm splice}$) using Hann windows
\begin{align}
\label{eq:splice}
w_{\rm off}(f) &= \frac{1}{2}\left[1+\cos\left(\frac{\pi [f - s_i]}{s_f - s_i}\right)\right],\\
w_{\rm on}(f) &= \frac{1}{2}\left[1-\cos\left(\frac{\pi [f - s_i]}{s_f - s_i}\right)\right].
\end{align}
The amplitude of the hybrid waveform is then
\begin{widetext}
\begin{equation}
|\tilde h_{\rm hybrid}(f)| =\
\left\{\begin{array}{lc}
|\tilde h_{\rm BBH}(f)|, & \, f \le s_i \\
w_{\rm off}(f) |\tilde h_{\rm BBH}(f)| + w_{\rm on}(f) |\tilde h_{\rm NR}(f)|, & \, s_i < f \le s_f \\
|\tilde h_{\rm NR}(f)|, & \, f>s_f,
\end{array}\right.
\end{equation}
and the phase is
\begin{equation}
\label{eq:hybridphase}
\Phi_{\rm hybrid}(f) =\
\left\{\begin{array}{lc}
\Phi_{\rm BBH+T}(f), & \, f \le s_i \\
w_{\rm off}(f) \Phi_{\rm BBH+T}(f) + w_{\rm on}(f) [\Phi_{\rm NR}(f) + 2\pi f \tau + \phi], & \, s_i < f \le s_f \\
\Phi_{\rm NR}(f) + 2\pi f \tau + \phi, & \, f>s_f.
\end{array}\right.
\end{equation}
\end{widetext}

\begin{figure*}[!htb]
\begin{center}
\includegraphics[width=2.9in]{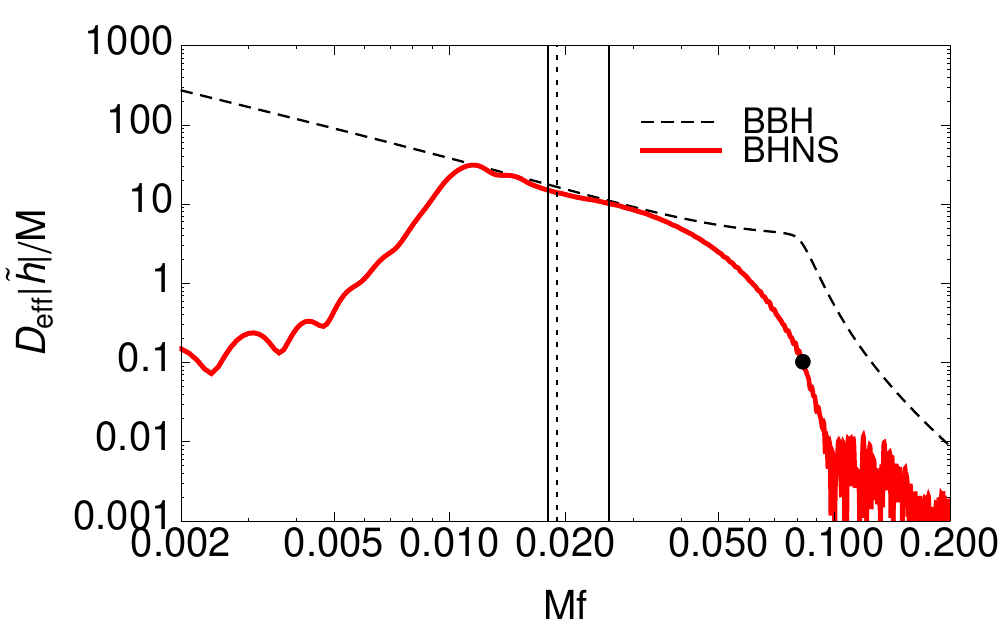}
\includegraphics[width=2.9in]{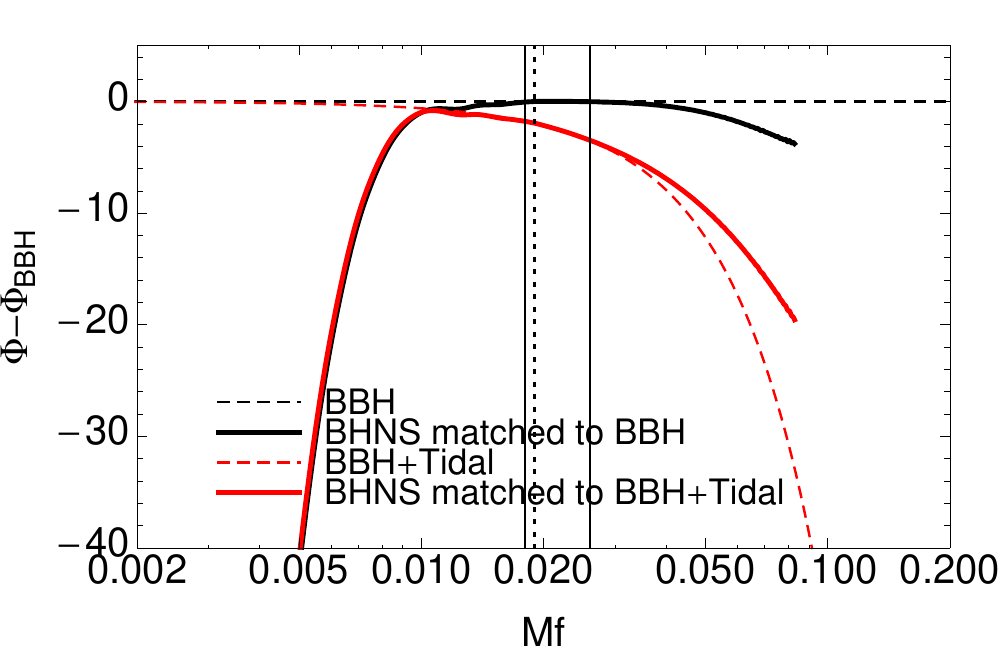}
\end{center}
\caption{ \label{fig:matching}
Amplitude $D_{\rm eff} |\tilde h(Mf)| / M$ (left) and phase $\Phi(Mf)$ (right) for a numerical BHNS waveform matched to the PhenomC BBH waveform with and without the tidal correction $\psi_T$ (Eq.~\eqref{eq:1pntaylorf2}). The waveform parameters are $(\chi_{\rm BH} = 0, Q = 2, M_{\rm NS} = 1.35~M_\odot, {\rm EOS} = p.5\Gamma3.0)$. The matching window $f_i<f<f_f$ is bounded by solid vertical lines, and the splicing window $s_i<f<s_f$, which begins at $s_i = f_i$, is bounded by dotted vertical lines. Note that matching the numerical BHNS waveform to a BBH waveform without tidal corrections, as was done in Paper~I, results in ignoring a phase term that accumulates linearly even after the matching region, and underestimates the effect of matter. We truncate the waveform when the amplitude drops below 0.1 denoted by a black dot.}
\end{figure*}

A hybrid waveform for the system ($\chi_{\rm BH} = 0, Q = 2, M_{\rm NS} = 1.35~M_\odot, \text{EOS} = p.5\Gamma 3.0$) is shown in Fig.~\ref{fig:matching}, where we matched the numerical waveform to the PhenomC waveform with and without the inspiral tidal correction $\psi_T$. In the right panel of Fig.~\ref{fig:matching} we show four phases relative to the PhenomC BBH phase $\Phi_{\rm BBH}$. The black dashed curve is the PhenomC BBH phase. The solid black curve is the numerical BHNS waveform phase after it is matched directly to the BBH waveform as was done in Paper~I\footnote{Paper~I used a time-domain method to match a numerical BHNS waveform to an EOB BBH inspiral waveform, and the time-domain matching interval used in Paper~I corresponds to a frequency interval slightly less than that shown here.}. The red dashed curve represents the phase of the tidally corrected inspiral waveform $\Phi_{\rm BBH+T}$, and the curve is given by the analytic expression for the inspiral tidal correction $\Phi_{\rm BBH+T}(Mf) - \Phi_{\rm BBH}(Mf) = \psi_T(Mf)$. Finally, the solid red curve is the numerical BHNS waveform matched to the tidally corrected inspiral waveform. We note that the difference after the matching window between the solid red and solid black curves depends only on the difference in the matching term $2\pi (Mf) (\tau/M)+\phi$ in the hybrid phase $\Phi_{\rm hybrid}$ (Eq.~\eqref{eq:hybridphase}) between when the inspiral tidal correction $\psi_T$ is (red curve) or is not (black curve) included. Because this term is linear, the difference between the two curves grows linearly. This difference is approximately $\psi_T(Mf_{\rm mid}) + (Mf - Mf_{\rm mid}) \psi_T' (Mf_{\rm mid})$, where $Mf_{\rm mid}$ is the midpoint of the matching interval and the $'$ indicates a derivative with respect to $Mf$. As we will see below, this linear term has a large impact on the measurability of tidal parameters.

\subsection{Sensitivity of hybrid waveform to matching parameters}

In the hybridization procedure described above, we are free to choose the window width $\Delta t_{\rm win}$ used in Eq.~\eqref{eq:window} as well as the matching interval $\Delta Mf_{\rm match} = Mf_f - Mf_i$ and midpoint of the matching interval $Mf_{\rm mid} = (Mf_i + Mf_f)/2$ used in Eq.~\eqref{eq:leastsquare}. If the numerical waveform were long and identical to the inspiral waveform within some interval, the choice of these free parameters would have no impact on the values of $\tau$ and $\phi$. However, there are several sources of error. Because the numerical waveform has finite length, the beginning of the waveform needs to be windowed before Fourier transforming the waveform to reduce the Gibbs phenomenon. The matching interval should exclude as much of the beginning of the waveform as possible because the numerical simulation takes time to settle down from inexact initial conditions which includes some initial eccentricity. It should also exclude the merger and ringdown which cannot be described by the tidal terms for the analytic inspiral waveform. On the other hand, the matching window must be wide enough to average over ringing from the Gibbs phenomenon that remains after windowing, the effects of eccentricity in the simulations, and other numerical noise. To isolate the effect that each of these waveform errors has on the values of $\tau$ and $\phi$, we will introduce them sequentially to the waveform that is matched to the inspiral waveform.

\begin{figure}[!htb]
\begin{center}
\includegraphics[width=3.2in]{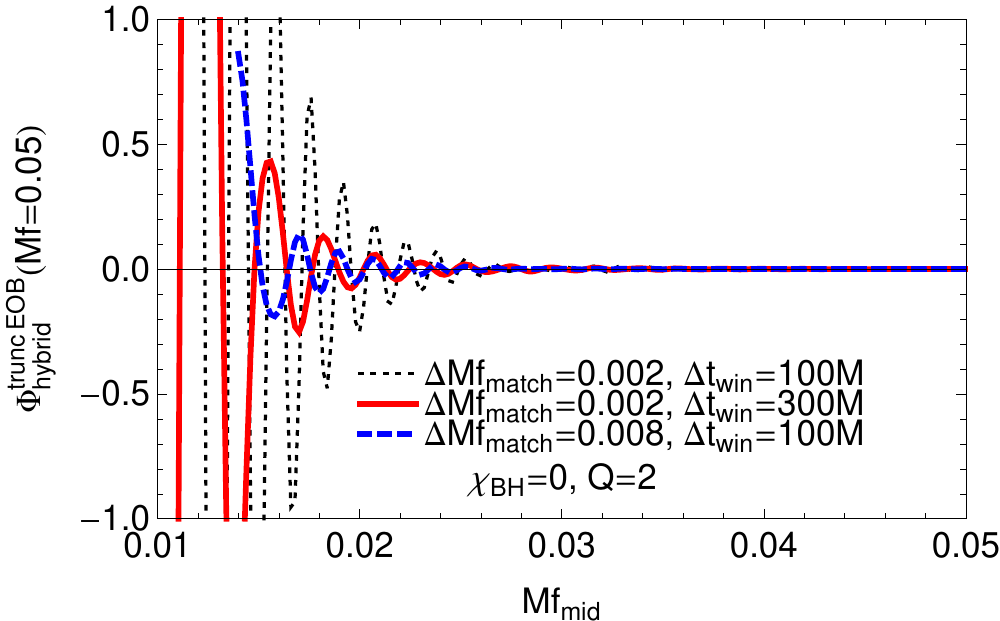}\\
\includegraphics[width=3.2in]{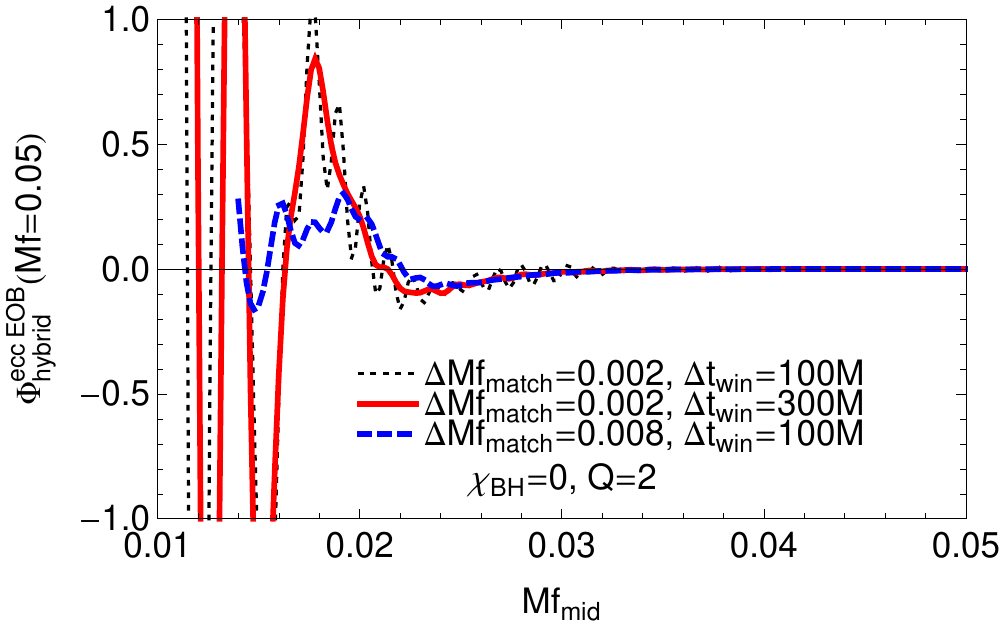}
\end{center}
\caption{ \label{fig:EOBmatching}
Dependence of hybrid waveform phase at $Mf = 0.05$ on window width $\Delta t_{\rm win}/M$ and matching interval with width $\Delta Mf_{\rm match}$ and midpoint $Mf_{\rm mid}$. Top panel: An EOB waveform, truncated to only include the last $800M$ before merger, is Fourier transformed then matched to the same EOB waveform that has not been truncated. The oscillations in the hybrid phase result from the Gibbs phenomenon. This can be partially suppressed by increasing the window width $\Delta t_{\rm win}/M$ as well as the width of the matching window $\Delta Mf_{\rm match}$. Bottom panel: The effect of eccentricity can be evaluated by matching an eccentric EOB waveform to a long, zero eccentricity EOB waveform. Here, an eccentric EOB waveform is generated by using the quasicircular initial conditions $M\Omega_0 = 0.028$. The dependence of the hybrid phase on the matching interval can be reduced by using a larger matching window $\Delta Mf_{\rm match}$. For both panels, $\chi_{\rm BH} = 0$ and $Q=2$. Because the overall phase of the hybrid waveform is arbitrary, we have set it to 0 in this figure when $Mf = 0.05$.}
\end{figure}

We first examine the effect of the Gibbs phenomenon, present in finite length waveforms, on the time and phase shifts $\tau$ and $\phi$. To do this, we begin with an (effectively) infinite length time-domain EOB waveform that includes the inspiral, merger, and ringdown. We then mimic a numerical waveform by making a truncated copy of this EOB waveform that starts $\sim 10$ GW cycles ($800M$) before merger. We window the first $\Delta t_{\rm win}$ and Fourier transform this truncated waveform, then match it to the Fourier transformed original waveform such that the only matching error is due to the Gibbs phenomenon. In the top panel of Fig.~\ref{fig:EOBmatching} we see the post-matching hybrid phase at $Mf=0.05$ depends on the window width $\Delta t_{\rm win}$ and the frequency interval defined by $\Delta Mf_{\rm match}$ and $Mf_{\rm mid}$. However, if we increase either the window width (from $\Delta t_{\rm win} = 100M$ to $300M$) or match over a larger frequency interval (from $\Delta Mf_{\rm match} = 0.002$ to $0.008$), we can reduce the dependence of the hybrid phase on the Gibbs Phenomenon.  

We next consider the effect of eccentricity on the hybrid phase in the bottom panel of Fig.~\ref{fig:EOBmatching}. The numerical BHNS simulations begin with quasicircular (zero radial velocity) initial conditions that ignore the small radial velocity due to radiation reaction. As a result, the inexact initial conditions lead to a small initial eccentricity ($e_0\sim0.03$), which eventually dies down after several orbits. We can mimick this effect by generating EOB waveforms with equivalent eccentricity by starting the EOB equations of motion with the same quasicircular (zero radial velocity) initial conditions as the simulations found by ignoring the radiation reaction term in the EOB equations of motion~\cite{BuonannoDamour2000}. We match an EOB waveform with the quasicircular initial conditions $M\Omega_0 = 0.028$ to an effectively infinite length, zero eccentricity EOB  waveform with otherwise identical parameters. As with the top panel, the eccentric EOB waveform exhibits Gibbs oscillations because it has a finite length, and this effect can be reduced by increasing the window width and frequency matching interval. There is also an additional offset that results from the initial eccentricity, and this offset eventually dies down around $Mf_{\rm mid} \sim 0.03$.

\begin{figure}[!htb]
\begin{center}
\includegraphics[width=3.2in]{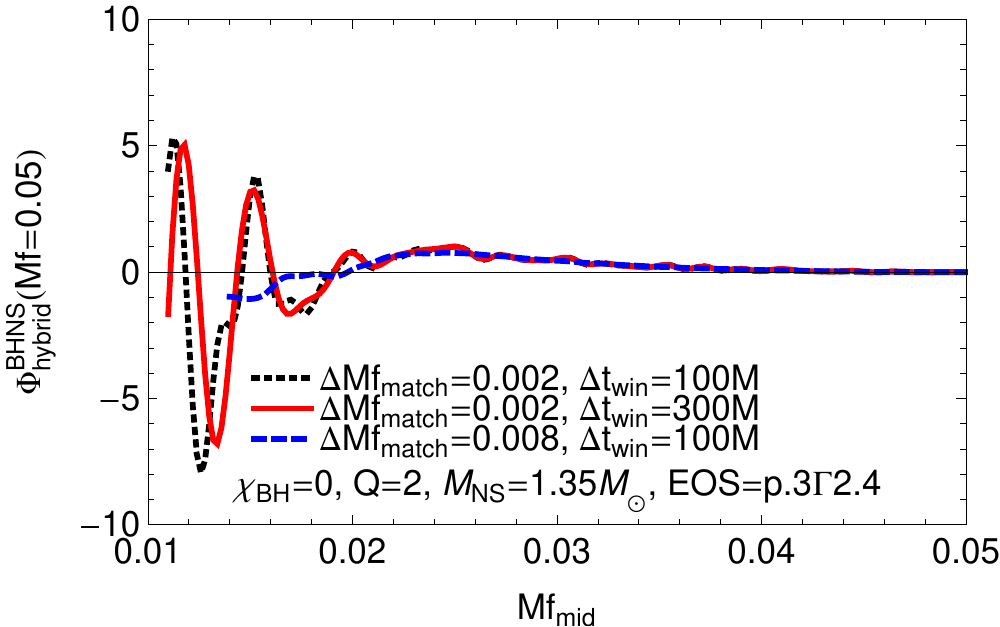}\\
\includegraphics[width=3.2in]{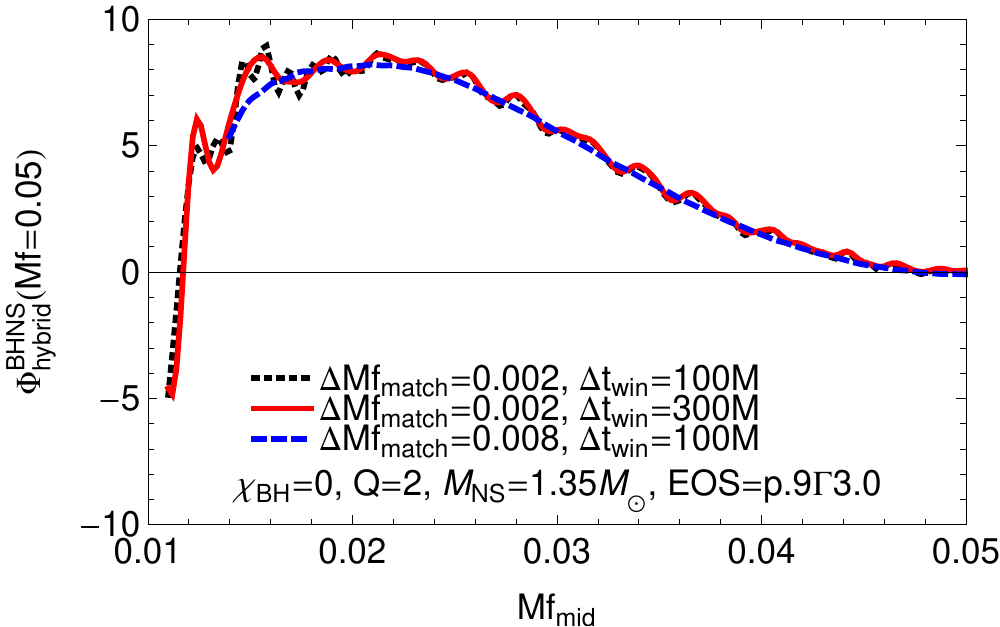}\\
\includegraphics[width=3.2in]{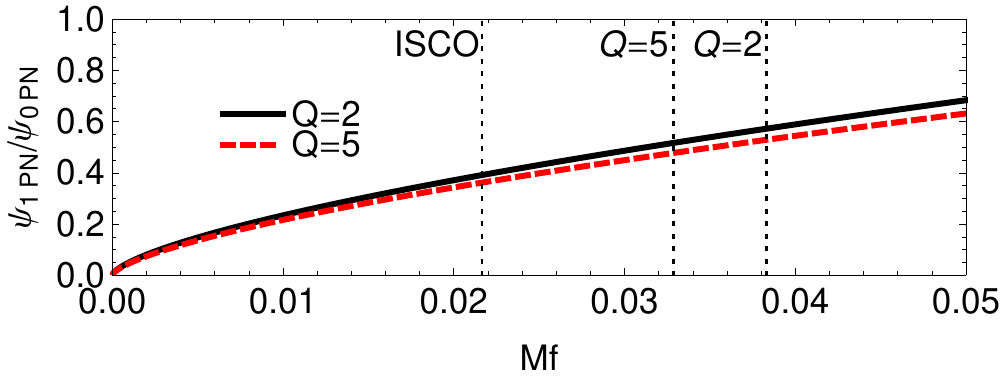}\\
\end{center}
\caption{ \label{fig:BHNSmatching}
Dependence of hybrid waveform phase at $Mf = 0.05$ on window width $\Delta t_{\rm win}/M$ and matching interval as in Fig.~\ref{fig:EOBmatching}. Top panel: BHNS simulation with parameters \{$\chi_{\rm BH} = 0$, $Q = 2$, $M_{\rm NS} = 1.35M_\odot$, $\mathrm{EOS} = p.3 \Gamma 2.4$\} is matched to the PhenomC BBH waveform with tidal phase corrections. Middle panel: BHNS simulation with parameters \{$\chi_{\rm BH} = 0$, $Q = 2$, $M_{\rm NS} = 1.35M_\odot$, $\mathrm{EOS} = p.9 \Gamma 3.0$\}. Bottom panel: Relative contribution of the 1PN tidal phase correction to the leading order tidal phase correction provides a crude estimate of the error in the tidal phase correction. For reference, dashed vertical lines give the ISCO frequency $1/(6^{3/2}\pi)$ as well as the innermost circular orbit for nonrotating black holes, defined by the minimum of the 3PN energy (Eq.~(194) of~\cite{Blanchet2006Review}), for mass ratios of $Q = 2$ and 5.
}
\end{figure}

Finally, in Fig.~\ref{fig:BHNSmatching} we match two numerical BHNS simulations to PhenomC inspiral waveforms with tidal corrections to generate a full BHNS IMR hybrid. If the inspiral waveform and numerical waveform are identical within a frequency interval, then the hybrid phase will be independent of the matching region within that interval, and there will therefore be a plateau in the curve in Fig.~\ref{fig:BHNSmatching}. We therefore identify the best matching region as the region centered on the maxima $Mf_{\rm mid} \approx 0.022$ in the top panel and $Mf_{\rm mid} \approx 0.020$ in the bottom panel. In addition, as in Fig.~\ref{fig:EOBmatching}, increasing the window width and matching interval reduces oscillations due to Gibbs phenomena. 

We note that the best matching region in Fig.~\ref{fig:BHNSmatching} overlaps somewhat with frequencies where eccentricity still effects the hybrid phase as seen in Fig.~\ref{fig:EOBmatching}. We can move the matching region to slightly higher frequencies; however, the analytic tidal correction will rapidly become inaccurate. We can estimate roughly the error of the inspiral tidal phase term in the bottom panel of Fig.~\ref{fig:BHNSmatching} as the ratio of the 1PN to leading tidal corrections (defined by $\psi_{1PN}/\psi_{0PN} = (a_1/a_0)(\pi M f)^{2/3}$).

Although the best choice of matching frequencies vary slightly depending on $\chi_{\rm BH}$, $Q$, and the EOS, setting the matching region to be the same for all waveforms changes the final hybrid phase no more than the matching uncertainties described above. From the above considerations, we choose as our windowing width $\Delta t_{\rm win} = 300M$, and matching parameters $\Delta Mf_{\rm match} = 0.008$ and $Mf_{\rm mid} = 0.022$ such that the start and end matching frequencies are $Mf_i = 0.018$ and $Mf_f = 0.026$. In the matching region, the amplitude and phase of the inspiral and shifted numerical waveforms agree reasonably well, so the choice of $s_i$ and $s_f$ does not significantly effect the results. We choose $Ms_i = Mf_i$ and $Ms_f = Ms_i + 0.001$.

\section{Parameter estimation}
\label{sec:4}
\label{sec:estimation}

The primary goal of this paper is to determine how accurately EOS parameters can be measured from BHNS observations. In this section we will discuss the statistical error associated with detector noise as well as the systematic error that results from using inexact waveform templates to estimate parameters. In the next three sections, we will then determine the combination of EOS parameters that is best measured and generate an analytic waveform model matched to numerical waveforms that we then use to estimate the statistical and systematic errors in measuring that EOS parameter combination.

The output of a gravitational-wave detector $s(t; \vec\theta_T) = n(t) + h_E(t; \vec\theta_T)$ is the sum of detector noise $n(t)$ and a possible gravitational-wave signal exactly described by $h_E(t; \vec\theta_T)$ with the true parameters $\vec\theta_T$. We assume the noise is a stationary, Gaussian time series, and therefore characterized by its power spectral density (PSD) $S_n(|f|)$ defined by the ensemble average
\begin{equation}
\langle \tilde{n}(f) \tilde{n}^\ast(f') \rangle = \frac{1}{2} \delta(f-f') S_n(|f|),
\end{equation}
and its probability distribution
\begin{equation}
p_n[n(t)] \propto e^{-(n, n)/2}.
\end{equation}
Here, $(a, b)$ is the usual inner product between two time series $a(t)$ and $b(t)$ weighted by the PSD
\begin{equation}
(a, b)=4{\rm Re} \int_0^\infty \frac{\tilde a(f) \tilde b^*(f)}{S_n(f)}\,df.
\end{equation}
The gravitational wave signal is given in terms of the two polarizations of the gravitational wave by
\begin{equation}
h_E(t; \vec\theta_T) = F_+ h_{E+}(t; \vec\theta_T) + F_\times h_{E\times}(t; \vec\theta_T),
\end{equation}
where $F_{+,\times}$ are the detector response functions and depend on the location of the binary and the polarization angle of the waves.  As in Paper I, we assume the binary is optimally located at the zenith of the detector and optimally oriented with its orbital axis along the line of sight.  This condition is equivalent to averaging $h_+$ and $h_\times$ ($F_+ = F_\times = 1/2$).

In searches for gravitational-wave signals from compact binary mergers, a set of templates $h(t; \vec\theta)$ with parameters $\vec\theta$ are compared to the signal $s(t)$. The parameters that maximize the signal to noise ratio (SNR)
\begin{equation}
\rho = \frac{(h, s)}{\sqrt{(h, h)}}
\end{equation}
are the best estimate of the true parameters $\vec\theta_T$. For a template $h_E(t; \vec\theta)$ that exactly represents the true waveform, we will denote the best estimate of the true parameters produced by the exact template by $\vec\theta_E$. In practice, however, we only have an approximate template $h_A(t; \vec\theta)$, and we will denote the best estimate produced by this approximate template by $\vec\theta_A$.

In the large SNR limit, the difference $\Delta\vec\theta = \vec\theta_A - \vec\theta_T$ between the best estimate using an approximate template and the true parameters of the binary system obeys a Gaussian distribution~\cite{FinnChernoff1993}. Specifically, for $N$ parameters, the conditional probability of the error $\Delta\vec\theta$ given the best estimate $\vec\theta_A$ is
\begin{equation}
p(\Delta\vec\theta | \vec\theta_A) = \frac{1}{\sqrt{(2\pi)^N \mathrm{det}(\Sigma_{ij})}} e^{-\frac{1}{2} \Sigma^{-1}_{ij} (\Delta\theta_i - \langle\Delta\theta_i\rangle )( \Delta\theta_j - \langle\Delta\theta_j\rangle)}.
\end{equation}
The mean is found to be approximately~\cite{CutlerVallisneri2007, CreightonAnderson2011}
\begin{equation}
\langle\Delta\theta_i\rangle \approx - \Gamma^{-1}_{ij} ( \delta h(\vec\theta_A), \partial_j h_A(\vec\theta_A) ),
\end{equation}
where
\begin{equation}
\Gamma_{ij} = (\partial_i h_A(\vec\theta_A), \partial_j h_A(\vec\theta_A))
\end{equation}
is the Fisher matrix and $\delta h(t; \vec\theta) = h_A(t; \vec\theta) - h_E(t; \vec\theta)$ is the difference between the approximate and exact waveform templates. The covariance between the parameters is~\cite{FinnChernoff1993}
\begin{equation}
\Sigma_{ij} \equiv \langle (\Delta\theta_i - \langle\Delta\theta_i\rangle )( \Delta\theta_j - \langle\Delta\theta_j\rangle) \rangle = \Gamma^{-1}_{ij},
\end{equation}
and the variance in $\Delta\theta_i$ is therefore
\begin{equation}
\label{eq:sigma}
\sigma_i^2 \equiv \langle ( \Delta\theta_i - \langle \Delta\theta_i \rangle )^2 \rangle = \Gamma^{-1}_{ii},
\end{equation}
where the repeated indices in $\Gamma^{-1}_{ii}$ do not represent summation. The statistical error ellipsoid to $n$ standard deviations is a contour of $p(\Delta\vec\theta | \vec\theta_A)$ given by
\begin{equation}
( \Delta\theta_i - \langle\Delta\theta_i\rangle )( \Delta\theta_j - \langle\Delta\theta_j\rangle ) \Sigma^{-1}_{ij} = n^2.
\end{equation}
In addition, we identify the quantity $\Delta\vec\theta_{\rm syst} \equiv \langle\Delta\vec\theta\rangle$ as the systematic error that results from using an approximate waveform template instead of the exact waveform template. We will use this expression below as a criteria for the accuracy of our analytic BHNS waveform model.

\section{Best measured EOS parameter}
\label{sec:5}

In Paper I we found that, during the merger and ringdown, the best-measured combination of EOS parameters for nonspinning BHNS systems was consistent with the tidal deformability $\Lambda$. We used hybrid waveforms that ignored the inspiral tidal correction $\psi_T$ (Eq.~\eqref{eq:1pntaylorf2}), and only included EOS information from the merger and ringdown of the numerical part of the waveform. We then evaluated a restricted two-parameter Fisher matrix for the EOS parameters $\log(p_1)$ and $\Gamma$, ignoring possible correlations between the EOS parameters and the other parameters. In this section we compare the measurability of EOS parameters for only the merger and ringdown to a waveform that includes EOS information in the full IMR hybrid waveform, and we do this for three combinations of mass ratio and black hole spin. In the next two sections we will address the issue of correlations between EOS and non-EOS parameters by constructing an analytic BHNS waveform and calculating the complete Fisher matrix.

As in Paper I, we evaluate the Fisher matrix from a set of hybrid waveforms by differentiating the waveform with respect to each parameter using finite differencing with two or more waveforms for each parameter. We follow the third method in Appendix~A of Paper I which results in the greatest accuracy given the phase difference between waveforms which can be several radians for EOS parameters. Specifically, we decompose each Fourier transformed hybrid waveform into the log of the amplitude $\ln A(f; \vec\theta)$ and accumulated phase $\Phi(f; \vec\theta)$
\begin{equation}
\label{eq:logampphase}
\tilde h(f; \vec\theta) = e^{\ln A(f; \vec\theta) + i \Phi(f; \vec\theta)},
\end{equation}
then evaluate $\partial_i \ln A$ and $\partial_i \Phi$ individually. The derivative is now approximated by
\begin{equation}
\partial_i \tilde h(f; \vec\theta) \approx e^{\ln A(f; \vec\theta) + i \Phi(f; \vec\theta)}\left( \frac{\Delta\ln A(f; \vec\theta)}{\Delta\theta^i} + i \frac{\Delta \Phi(f; \vec\theta)}{\Delta\theta^i} \right),
\end{equation}
where $\Delta/\Delta\theta^i$ represents central differencing, and $\ln A$ and $\Phi$ are evaluated at the midpoint with linear interpolation.

Calculating the complete Fisher matrix using hybrid waveforms requires one to evaluate partial derivatives with respect to all parameters at a single point. For an aligned-spin BHNS system with 2 EOS parameters and a single detector, the waveform will have the form
\begin{equation}
\begin{split}
\tilde h(f; \vec\theta) &= \frac{1}{D_{\rm eff}} g_A(f; \mathcal{M}, \eta, \chi_{\rm BH}, \log(p_1), \Gamma) \\
& \times e^{i [ 2\pi f t_c + \phi_c + g_\Phi(f; \mathcal{M}, \eta, \chi_{\rm BH}, \log(p_1), \Gamma) ] },
\end{split}
\end{equation}
where $g_A$ and $g_\Phi$ are generic functions, and there are 8 parameters. The 5 intrinsic parameters are the chirp mass $\mathcal{M} = (M_{\rm BH} M_{\rm NS})^{3/5}/M^{1/5}$, symmetric mass ratio $\eta = M_{\rm BH} M_{\rm NS}/M^2$, black hole spin $\chi_{\rm BH}$\footnote{There will also be a neutron-star spin contribution $\chi_{\rm NS}$. Magnetic dipole radiation, however, is expected to spin down a NS to a small fraction of the Kepler frequency well before the binary reaches the detector band. Furthermore, for the PhenomC waveform, the aligned spins of the two bodies are approximated by the single mass-weighted average-spin parameter $\chi_{\rm avg}$ (Eq.~\eqref{eq:chiavg}). Since we will use $\chi_{\rm BH}$ as our only spin parameter it effectively becomes a linear combination of the BH and NS spins. Incorporating NS with significant spins would require a different inspiral waveform model with two separate spin parameters as well as BHNS simulations with spining NS.}, and the 2 EOS parameters $\log(p_1)$ and $\Gamma$. The 3 extrinsic parameters, which can be differentiated analytically, are time of coalescence $t_c$, phase of coalescence $\phi_c$, and an effective distance $D_{\rm eff}$ that incorporates the true distance $D$ as well as the orientation and sky location of the binary. (For an optimally oriented and located binary, $D_{\rm eff} = D$). If using central differencing, this requires 10 hybrid waveforms for the 5 numerical derivatives at each point in the waveform parameter space, and is computationally expensive if one wants to explore the entire parameter space. In addition, in contrast to the small EOS dependent effects, small changes in $\mathcal{M}$, $\eta$, and $\chi_{\rm BH}$ can result in a large change in the phase of the waveform. This means that the simulations must be closely spaced in parameter space in order to accurately calculate derivatives, requiring a very large number of waveforms. In this section, we will therefore restrict the Fisher matrix calculation to the two EOS parameters $\log(p_1)$ and $\Gamma$, and will leave to the next section a better way to differentiate the other parameters.

\begin{figure}[!htb]
\begin{center}
\includegraphics[width=80mm]{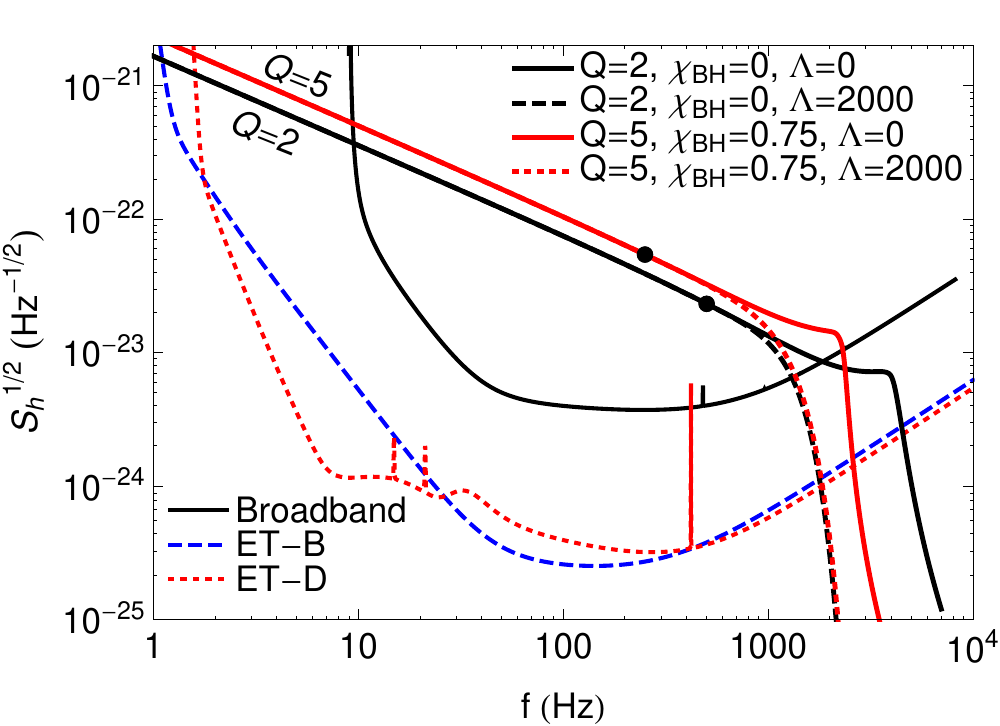}
\end{center}
\caption{ \label{fig:noisecurves}
(Color online) Noise PSD for broadband aLIGO, ET-B, and ET-D. Also shown are the weighted amplitudes $2 f^{1/2} |\tilde h|$ of phenomenological waveforms for 4 parameter values. For all waveforms, $M_{\rm NS} = 1.35M_\odot$ and $D_{\rm eff} = 100$~Mpc. Black circles represent the start of the waveform fit at $Mf = 0.01$.
}
\end{figure}

For the BHNS systems discussed here, the greatest departure from BBH behavior occurs for gravitational-wave frequencies in the range 300--3000~Hz. As a result, detector configurations optimized for detection of BHNS systems with low noise in the region below 300~Hz may not optimally estimate EOS parameters.  We therefore present results for the broadband aLIGO noise curve~\cite{LIGOnoise} and the ET-D noise curve~\cite{ETDnoise} shown in Fig.~\ref{fig:noisecurves}.  The broadband aLIGO configuration uses zero-detuning of the signal recycling mirror and a high laser power, resulting in significantly lower noise above 300~Hz at the expense of slightly higher noise at lower frequencies.  Several configurations have been considered for the Einstein Telescope denoted ET-B~\cite{ETBnoise}, ET-C~\cite{ETCnoise}, and ET-D~\cite{ETDnoise}.  We use the most recent ET-D configuration and note that in the 300--3000~Hz range all of the ET configurations have a similar sensitivity.  The published noise curves, and those used in this paper, are for a single interferometer of 10~km with a 90$^\circ$ opening angle.  The current ET proposal is to have three individual interferometers each with a 60$^\circ$ opening angle configured in an equilateral triangle.  This will shift the noise curve down appoximately 20\%~\cite{ETDnoise}.

\begin{figure}[!htb]
\begin{center}
\includegraphics[width=2.8in]{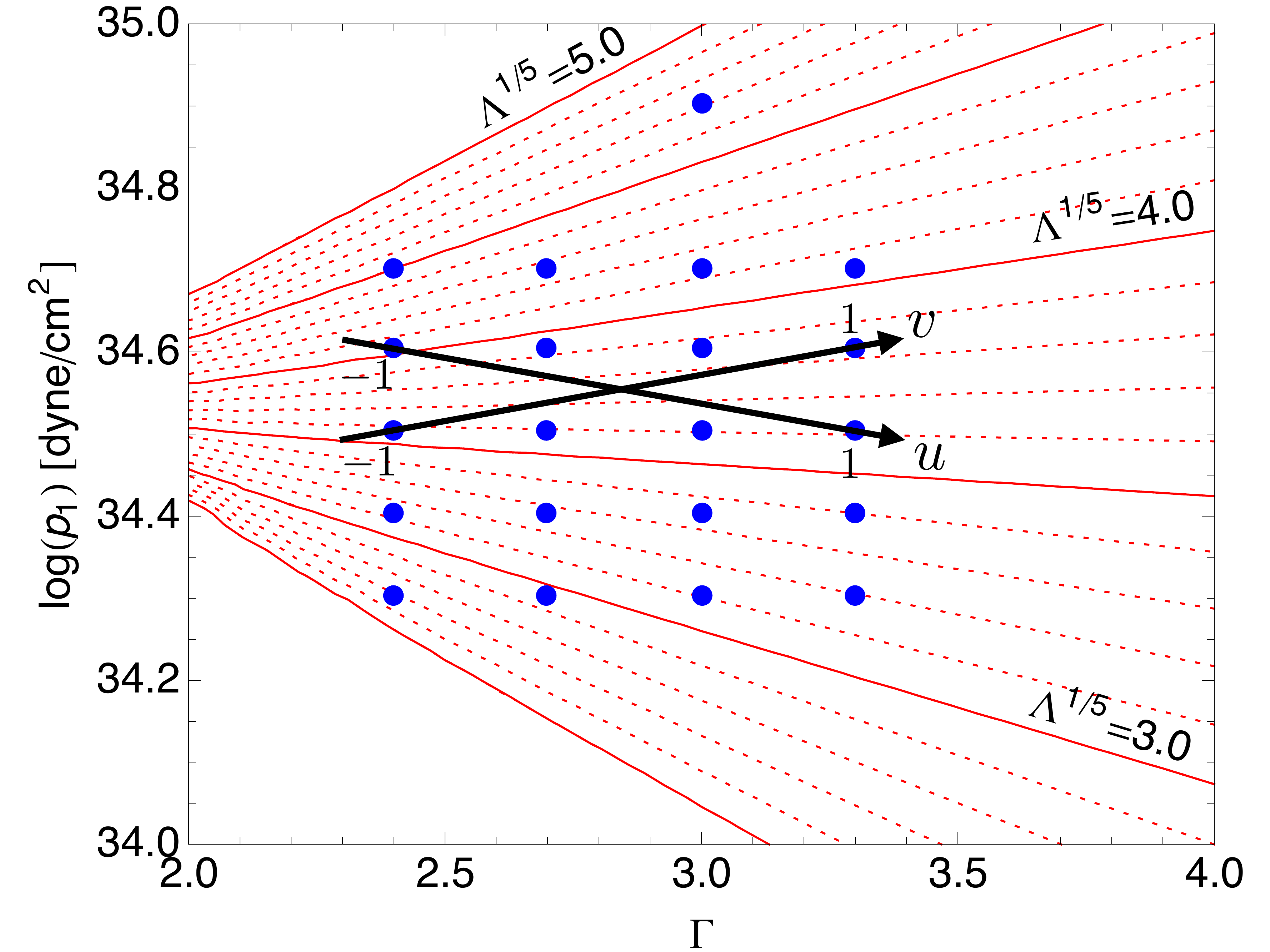}
\end{center}
\caption{ 
\label{fig:schematic}
Example of coordinates in the $u$--$v$ coordinate system used to compute derivatives in Eqs.~\eqref{eq:dhdlogp1} and~\eqref{eq:dhdgamma}. The $u$--$v$ coordinate system is chosen so the axes are not aligned with contours of $\Lambda^{1/5}$.
}
\end{figure}

The 1-$\sigma$ uncertainty ellipses in the EOS parameter space $\vec\theta = \{ \log(p_1), \Gamma \}$ are $\Delta\theta^i\Delta\theta^j\Gamma_{ij} = 1$. When calculating these 2-parameter error ellipses with the Fisher matrix using finite differencing, we sometimes find that two waveforms with the same $\log(p_1)$ but different $\Gamma$ are nearly identical, leading to derivatives that are dominated by numerical errors in the waveforms. To avoid this problem, we transform the $\Gamma$--$\log(p_1)$ coordinate system to the $u$--$v$ coordinate system shown in Fig.~\ref{fig:schematic}. We then evaluate the derivatives in this new coordinate system, and finally transform back with the chain rule
\begin{align}
\label{eq:dhdlogp1}
\frac{\partial\tilde h}{\partial\log(p_1)} &= \frac{\partial u}{\partial\log(p_1)} \frac{\partial\tilde h}{\partial u} + \frac{\partial v}{\partial\log(p_1)} \frac{\partial\tilde h}{\partial v}, \\
\label{eq:dhdgamma}
\frac{\partial\tilde h}{\partial\Gamma} &= \frac{\partial u}{\partial\Gamma} \frac{\partial\tilde h}{\partial u} + \frac{\partial v}{\partial\Gamma} \frac{\partial\tilde h}{\partial v}.
\end{align}

\begin{figure}[!htb]
\begin{center}
\includegraphics[width=2.8in]{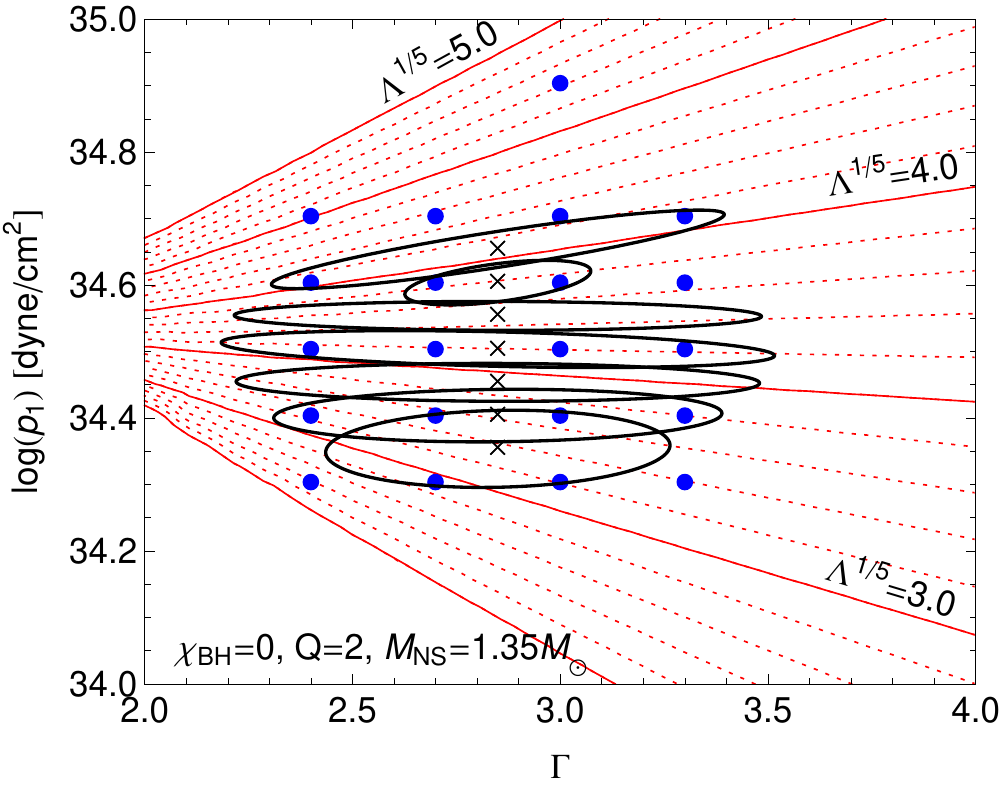}\\
\includegraphics[width=2.8in]{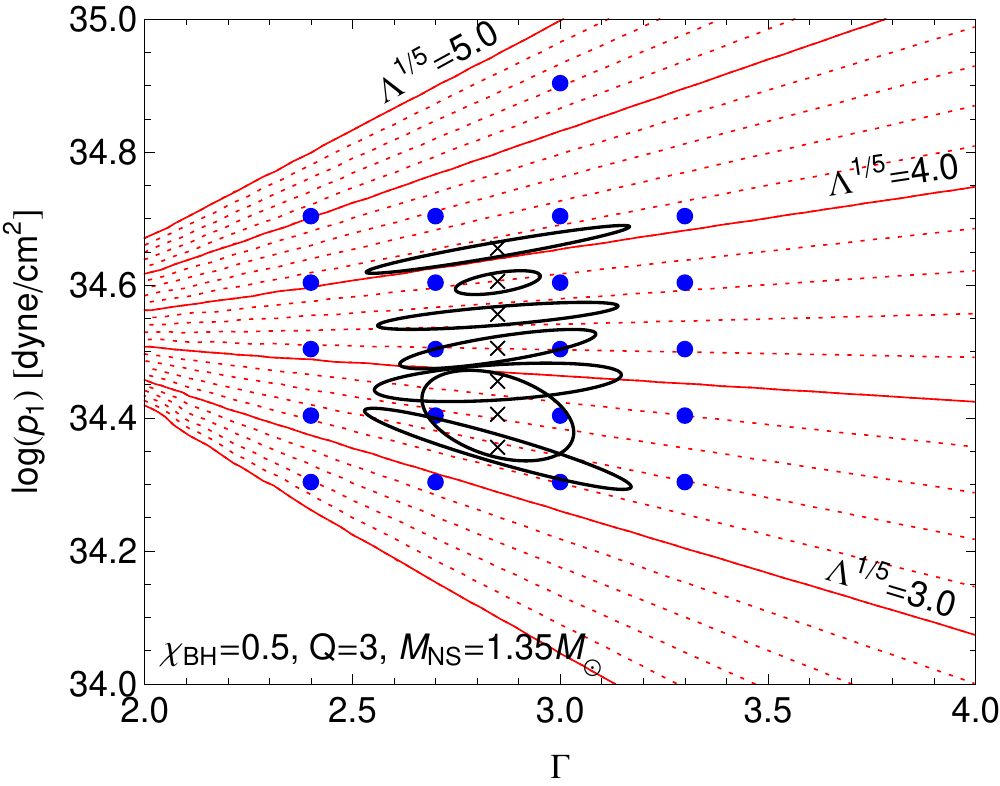}\\
\includegraphics[width=2.8in]{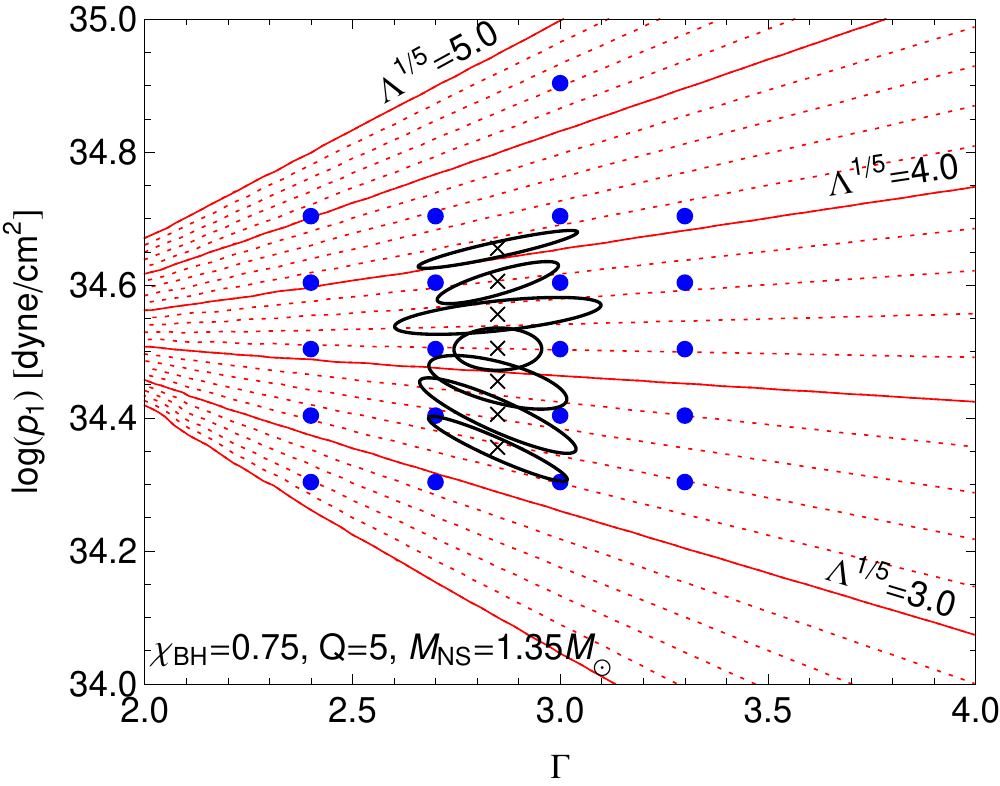}
\end{center}
\caption[Uncorrelated uncertainty in $\Lambda$ for merger-ringdown]{ 
\label{fig:ellipseETMR}
1-$\sigma$ error ellipses for the 2-parameter Fisher matrix using the ET-D noise curve. We use hybrid waveforms where the numerical BHNS waveform is matched to a PhenomC BBH inspiral waveform with no tidal correction. Binary is optimally oriented and at a distance of 100~Mpc. Values of $\chi_{\rm BH}$, $Q$, and $M_{\rm NS}$ are listed in each panel. Evenly spaced contours of constant $\Lambda^{1/5}$ are also shown.  Each ellipse is centered on the estimated parameter $\vec\theta_A$ denoted by a $\times$. Top: Matching window has width $\Delta Mf_{\rm match} = 0.008$ and is centered on $Mf_{\rm mid} = 0.016$. Middle: $\Delta Mf_{\rm match} = 0.008$ and $Mf_{\rm mid} = 0.016$. Bottom: $\Delta Mf_{\rm match} = 0.008$ and $Mf_{\rm mid} = 0.020$.  
}
\end{figure}

These ellipses are shown in Fig.~\ref{fig:ellipseETMR} for the ET-D noise PSD for both nonspinning and spinning simulations when the BHNS waveform is matched to a PhenomC BBH inspiral waveform with no tidal correction $\psi_T$. As in Paper~I we find the uncertainty contours are approximately aligned with tidal deformability contours $\Lambda^{1/5}$, and this holds for systems with spinning black holes as well. As in Paper~I we plot $\Lambda^{1/5}$ instead of $\Lambda$ because it is more closely related to the NS radius. We also note that the error ellipses found here for ($\chi_{\rm BH}=0$, $Q=2$, $M_{\rm NS}=1.35M_\odot$) using the PhenomC BBH inspiral waveform and frequency-domain match are very similar to the results found in Fig.~11 of Paper~I where we used the EOB BBH inspiral waveform and a time-domain match.

\begin{figure}[!htb]
\begin{center}
\includegraphics[width=2.8in]{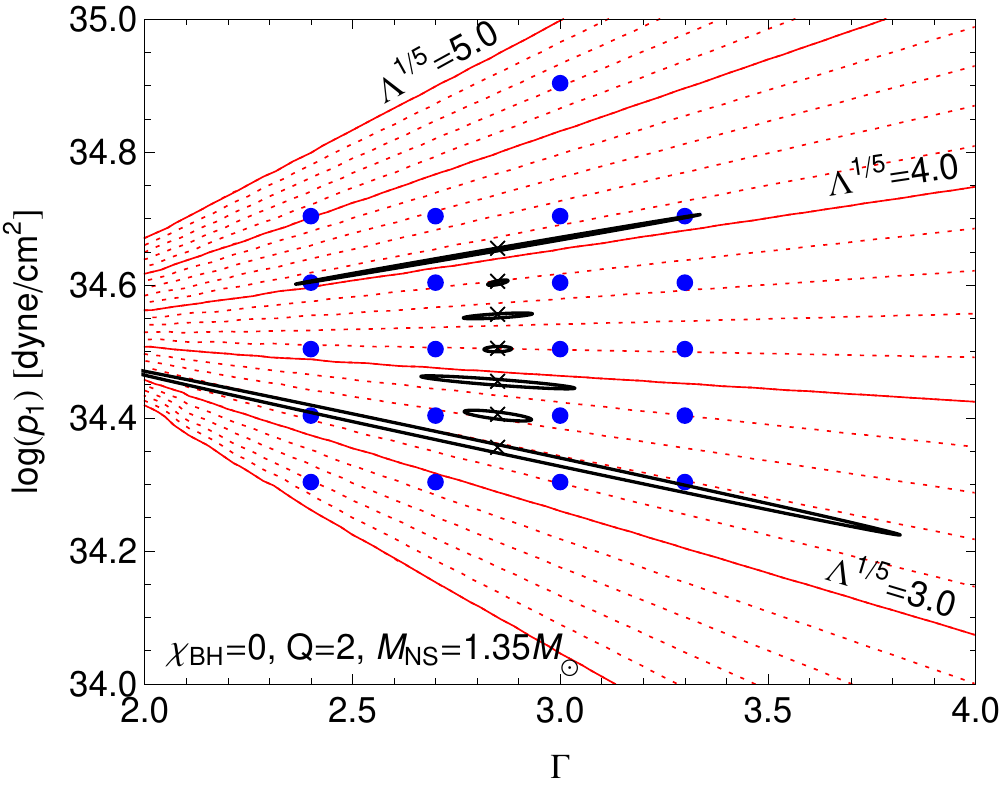}\\
\includegraphics[width=2.8in]{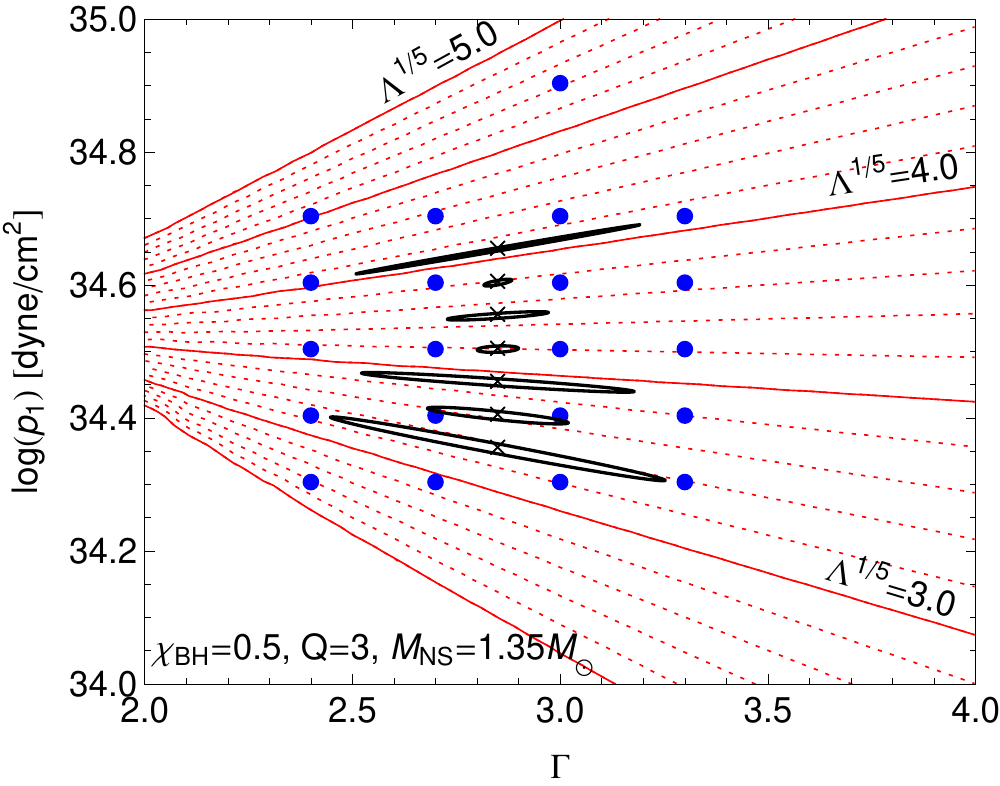}\\
\includegraphics[width=2.8in]{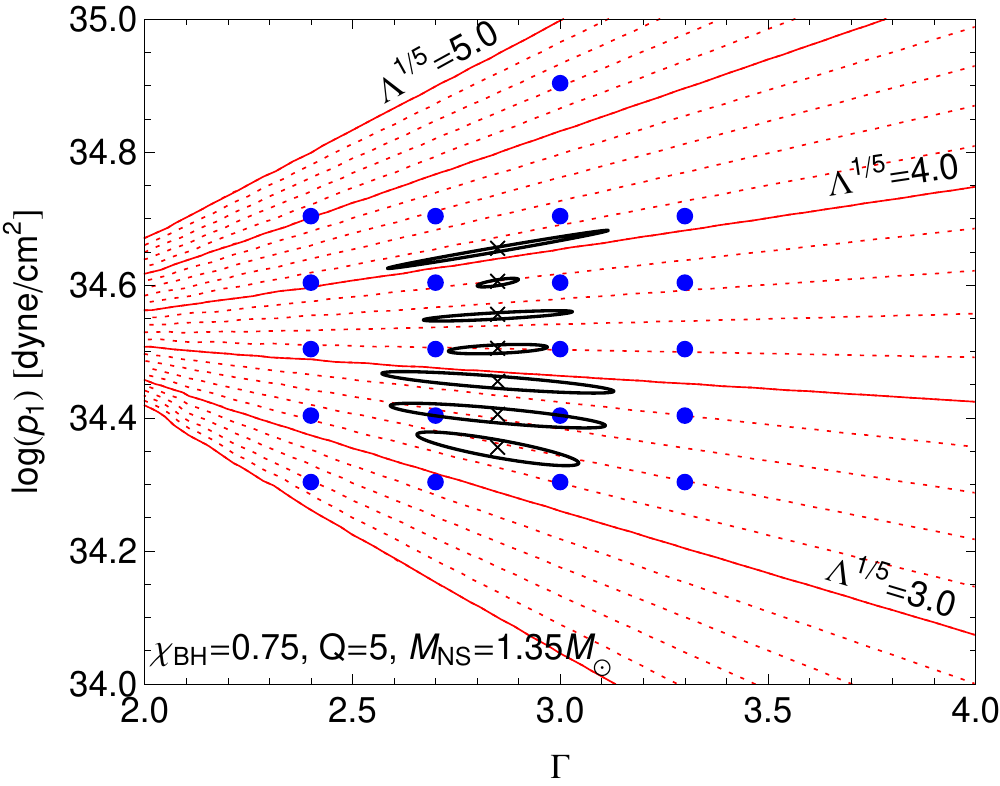}
\end{center}
\caption[Uncorrelated uncertainty in $\Lambda$ for inspiral-merger-ringdown]{ 
\label{fig:ellipseETIMR}
Same as Fig.~\ref{fig:ellipseETMR}, except the hybrid waveforms are generated by matching the numerical BHNS waveforms to PhenomC inspiral waveforms with the tidal correction $\psi_T$. For all panels, the matching window has width $\Delta Mf_{\rm match} = 0.008$ and is centered on $Mf_{\rm mid} = 0.022$.
}
\end{figure}

In contrast, when the tidal correction $\psi_T$ is added to the PhenomC inspiral waveform before generating a hybrid waveform, there is an improvement of roughly a factor of 3 in the measurability of $\Lambda$ as shown in Fig.~\ref{fig:ellipseETIMR}. The majority of the improvement arises because, as stated in the discussion of Fig.~\ref{fig:matching}, even though the inspiral tidal correction is small, the hybridization procedure also adds a tidal term that grows linearly with frequency to the merger and ringdown which is not present when the numerical waveform is joined to an inspiral waveform without tidal corrections. In addition, because the inspiral tidal correction $\psi_T$ and the tidal contribution to the matching term [$\psi_T(Mf_{\rm mid}) + (Mf - Mf_{\rm mid}) \psi_T' (Mf_{\rm mid})$] are analytically proportional to $\Lambda$, the ellipses align much more closely with the $\Lambda$ contours. We emphasize that the majority of the improvement comes from the above tidal contribution to the matching term, and not from the inspiral term $\psi_T$. As already noted~\cite{PannaraleRezzollaOhmeRead2011}, tidal interactions during the inspiral alone are not separately measurable. Finally we note that when using the broadband aLIGO noise curve instead of the ET-D noise curve, the error ellipses in Figs.~\ref{fig:ellipseETMR} and~\ref{fig:ellipseETIMR} have nearly identical shape and orientation but the size is a factor of $\sim 10$ larger.

We tentatively conclude that $\Lambda$ is the dominant EOS dependent quantity that can be measured for the merger and ringdown as well as for the inspiral. We have considered only the two-dimensional cross section $\Delta\vec\theta^{\rm non}_{\rm EOS} = 0$ of the eight-dimensional ellipsoid $\Delta\theta_i \Delta\theta_j \Gamma_{ij} = 1$, however, and have therefore ignored correlations between the EOS parameters and the non-EOS parameters.
 We partially address this deficiency in the next section by explicitly accounting for correlations between $\Lambda$ and the non-EOS parameters on a slice of constant $\Gamma$. One may still worry about the effect of correlations between $\Gamma$ and all other parameters.
Because the majority of the tidal contribution comes from the linearly growing matching term that analytically depends on $\Lambda$, we believe this assumption is mostly justified. We will test this assumption by calculating the systematic error in our phenomenological waveform model below which assumes $\Lambda$ is the EOS dependent parameter.

\section{Phenomenological BHNS waveform}
\label{sec:6}
\label{sec:phenomBHNS}

In Paper I and the section above, we assumed that  correlations between $\Lambda$ and the other parameters are negligible. To test this assumption, we must calculate the complete Fisher matrix for all parameters, and this requires us to evaluate partial derivatives with respect to all parameters at a single point. As discussed above, this is computationally difficult for hybrid waveforms using finite differencing. Another approach is to construct an analytic BHNS waveform model with free parameters that are fit to the hybridized numerical waveforms. This allows one to interpolate between the available simulations and to evaluate derivatives used in the Fisher matrix.

As found in Section~\ref{sec:5} (Figs.~\ref{fig:ellipseETMR} and~\ref{fig:ellipseETIMR}) and in Figs.~\ref{fig:hybridvarylambda}--\ref{fig:hybridvarylambda3}, a BHNS waveform is well approximated by a one-parameter deformation from a BBH waveform where $\Lambda=0$~\cite{BinningtonPoisson2009}. As shown in Figs.~\ref{fig:hybridvarylambda}--\ref{fig:hybridvarylambda3}, throughout the inspiral, merger, and ringdown, both the amplitude and phase of the Fourier transformed waveform monotonically decrease with respect to a BBH waveform as frequency increases and as $\Lambda$ increases\footnote{For mass ratios of $Q = 4$ and 5, the PhenomC amplitude is sometimes slightly less than the BHNS waveform with the softest EOS during ringdown around $Mf\sim 0.1$ as seen in Figs.~\ref{fig:hybridvarylambda} and~\ref{fig:hybridvarylambda3}. The amplitude of the corresponding EOB waveform, however, is always greater than the BHNS amplitudes.}. We thus write the BHNS waveform as a modification to a BBH waveform
\begin{equation}
\tilde h_{\rm BHNS}(Mf; \vec\theta) = \tilde h_{\rm BBH}(Mf; \vec\theta) r(Mf; \vec\theta) e^{i \Delta\Phi(Mf; \vec\theta)},
\end{equation}
where the ratio $r(Mf; \vec\theta) = |\tilde h_{\rm BHNS}(Mf; \vec\theta)| / |\tilde h_{\rm BBH}(Mf; \vec\theta)|$ is the amplitude correction, $\Delta\Phi(Mf; \vec\theta) = \Phi_{\rm BHNS}(Mf; \vec\theta) - \Phi_{\rm BBH}(Mf; \vec\theta)$ is a phase correction factor, and the 3 physical parameters that we will fit our waveforms to are $\vec\theta = \{ \eta, \chi_{\rm BH}, \Lambda \}$. We will then fit the quantities $r(Mf; \vec\theta)$ and $\Delta\Phi(Mf; \vec\theta)$ to the 134 hybrid waveforms listed in Table~\ref{tab:simulations}. 

For aligned-spin BBH systems, significant work has gone into developing analytic waveforms that include the complete inspiral, merger, and ringdown, and are calibrated to numerical BBH simulations. Unfortunately, there is no single BBH waveform model that has been calibrated for the ranges of mass ratio $Q \in [2, 5]$ and black hole spin $\chi_{\rm BH} \in [-0.5, 0.75]$ listed in Table~\ref{tab:simulations}. Furthermore, there are differences between the various analytic BBH waveforms for values of $Q$ and $\chi_{\rm BH}$ for which they have both been calibrated (See the PhenomC and EOB waveforms in Figs.~\ref{fig:hybridvarylambda}--\ref{fig:hybridvarylambda3}). Although small, these differences are still a non-negligible fraction of the difference between a BBH and BHNS waveform. As a result, we will have to separately calibrate our BHNS waveform model for each BBH model, and our waveform for very small values of $\Lambda$, where intrinsic errors in the analytic BBH models may dominate over tidal effects, will likely not be accurate. The two BBH models we use are the frequency-domain PhenomC~\cite{Santamaria2010} waveform and the time-domain EOB waveform~\cite{TaracchiniPanBuonanno2012} discussed in Section~\ref{sec:hybrid}.

\subsection{Fit based on PhenomC BBH approximation}

We fit the corrections to the PhenomC waveform, $r$ and $\Delta\Phi$, to our numerical BHNS waveforms. Although the PhenomC waveform is not calibrated using BBH waveforms with mass ratios of $Q=5$, we will fit $r$ and $\Delta\Phi$ to $Q=5$ BHNS waveforms anyway.

\subsubsection{Amplitude fit}

During the inspiral, because parameter estimation is much more sensitive to the fractional change in phase of the waveform than to the fractional change in amplitude, we ignore the very small amplitude correction from tidal interactions. During the merger and ringdown, however, amplitude corrections are important. We therefore write the amplitude correction as
\begin{equation}
\label{eq:r}
r(Mf; \vec\theta) =\
\left\{\begin{array}{lc}
1 & Mf \le Mf_A \\
e^{ - \eta\Lambda B(Mf; \vec\theta) } & Mf > Mf_A
\end{array}\right.,
\end{equation}
where $Mf_A$ is the boundary, chosen below, between the inspiral and merger for the amplitude fit. We have extracted the quantity $\eta \Lambda$ because, as $\eta \to 0$ (extreme mass ratio limit) or $\Lambda \to 0$ (no matter limit), the waveform should approach that of a BBH waveform. We now impose two requirements on the function $B(Mf; \vec\theta)$. (i) The amplitude must be continuous at the frequency $Mf_A$, so $B(Mf_A; \vec\theta) = 0$. (ii) Because the amplitude of the BHNS waveform is almost always less than that of the corresponding BBH waveform, we require $B(Mf; \vec\theta) \ge 0$ for $Mf \ge Mf_A$ and for all physical values of the parameters: $\eta \in [0, 0.25]$, $\chi_{\rm BH} \in [-1, 1]$, and $\Lambda \ge 0$.

Given the above restrictions, we find that a useful fitting function for the amplitude correction is $B(Mf; \vec\theta) = C (Mf - Mf_A)^D$, where $C$ and $D$ are free parameters, and $Mf_A = 0.01$. With this ansatz, we then do a nonlinear least-squares fit to determine the parameters $C$ and $D$. We find that over the 134 simulations, $D$ has a mean and standard deviation of $D \sim 3 \pm 0.5$, and because these parameters are highly correlated, we fix $D = 3$ so that $B = C(Mf - Mf_A)^3$. We then fit each waveform with the single parameter $C$. The parameter $C$ is then fit to the physical parameters. We find that for fixed $\eta$ and $\chi_{\rm BH}$, $C$ is approximately a linear function of $\Lambda$. We therefore use the function $C(\eta, \chi_{\rm BH}, \Lambda) = e^{b_0 + b_1\eta + b_2\chi_{\rm BH}} + \Lambda e^{c_0 + c_1 \eta + c_2 \chi_{\rm BH}}$, where the parameters $\{b_0, b_1, b_2, c_0, c_1, c_2\}$ are found with a nonlinear least-squares fit. We note that this function is positive for all physical values of the parameters $\eta$, $\chi_{\rm BH}$, and $\Lambda$. Given the small number of samples for $\eta$ and $\chi_{\rm BH}$ as well as the difficulty in extracting the small tidal contribution from numerical simulations, we have used as few parameters in our fit as possible rather than to over-fit noisy data with a large number of parameters. The final form of $B$ is
\begin{equation}
\label{eq:Bfit}
B(Mf; \vec\theta) = \left(e^{b_0 + b_1\eta + b_2\chi_{\rm BH}} + \Lambda e^{c_0 + c_1 \eta + c_2 \chi_{\rm BH}}\right)(Mf - Mf_A)^3,
\end{equation}
and the best-fit parameters are $\{b_0, b_1, b_2, c_0, c_1, c_2\} = \{-64.985, -2521.8, 555.17, -8.8093, 30.533, 0.64960\}$. Using this fit, we find typical fractional errors in $C$ of $\sim 30\%$ for $\Lambda \gtrsim 500$. However, errors can be significantly larger for $\Lambda \lesssim 500$ and $Q = 4$ and 5, where $r$ is small due to the small tidal interaction, and the error is dominated by numerical noise and uncertainty in the BBH waveform. This is not significant because, as we will find, the systematic error that results from poorly fitting $r$ is still smaller than the statistical error in $\Lambda$ for small values of $\Lambda$.

\subsubsection{Phase fit}

For the phase of the phenomenological waveform we choose the following ansatz
\begin{widetext}
\begin{equation}
\label{eq:deltaphi}
\Delta\Phi(Mf; \vec\theta) =\
\left\{\begin{array}{lc}
\psi_T(Mf; \vec\theta) & Mf \le Mf_\Phi \\
- \eta\Lambda E(Mf; \vec\theta) + \psi_T(Mf_\Phi; \vec\theta) + (Mf - Mf_\Phi) \psi_T' (Mf_\Phi; \vec\theta) & Mf > Mf_\Phi
\end{array}\right.,
\end{equation}
\end{widetext}
where $\psi_T$ is the frequency-domain tidal phase correction for the inspiral, and a $'$ denotes a derivative with respect to $Mf$. In this paper we will use the 1PN accurate TaylorF2 tidal correction (Eq.~\eqref{eq:1pntaylorf2}) for the inspiral. {This expression explicitly breaks the EOS-dependent contribution to the phase into three pieces: (i) the contribution due to the inspiral tidal correction $\psi_T(Mf;\vec\theta)$, (ii) the contribution due to the merger-ringdown dynamics $-\eta \Lambda E(Mf; \vec\theta)$ which we will fit to numerical simulations, and (iii) the term $\psi_T(Mf_\Phi; \vec\theta) + (Mf - Mf_\Phi) \psi_T' (Mf_\Phi; \vec\theta)$ that grows linearly after the transition frequency $Mf_\Phi$ and results from matching the phase and derivative of the merger to the tidally corrected inspiral as discussed in Section~\ref{sec:hybrid}.}

As in the amplitude fit, we have explicitly pulled out the quantity $\eta \Lambda$ in the first term $-\eta\Lambda E$ because the phase of the BHNS waveform should approach that of a BBH waveform as $\eta \to 0$ or $\Lambda \to 0$. We further require the remaining function $E(Mf; \vec\theta)$ to satisfy the following conditions: (i) $E(Mf_\Phi; \vec\theta) = 0$, (ii) $E'(Mf_\Phi; \vec\theta) = 0$, and (iii) $E(Mf; \vec\theta) \ge 0$ for $Mf \ge Mf_\Phi$ and for all physical values of $\eta$, $\chi_{\rm BH}$, and $\Lambda$. In this way, the function $E(Mf; \vec\theta)$ is determined fully by the numerical waveform and is independent of the inspiral tidal term $\psi_T$.

{A key feature of this ansatz for the phenomenological waveform is that, if we do not change the hybridization matching window $(Mf_i, Mf_f)$, an improved inspiral tidal phase term $\psi_T$ can be swapped in to Eq.~\eqref{eq:deltaphi} without requiring one to redo the following fit for $E(Mf; \vec\theta)$.  This is useful for estimating how an improved inspiral tidal correction effects the measurability of tidal parameters for the complete IMR waveform. We note, however, that an improved inspiral tidal term will lead to a slightly different optimal matching window $(Mf_i, Mf_f)$ for the hybridization procedure, and using the optimal matching window for the hybridization with the improved inspiral tidal term will require one to redo the fit for $E$.}

We find that each waveform can be accurately fit with a function of the form $E = G(Mf - Mf_\Phi)^H$, where $G$ and $H$ are free parameters, and unlike the amplitude fit where $Mf_A = 0.01$, we choose $Mf_\Phi = 0.02$ for the transition frequency because it is close to the midpoint ($Mf_{\rm mid} = 0.022$) of the hybrid matching interval. For the 134 BHNS waveforms the best fit for the parameter $H$ has a relatively narrow range of $\sim 2 \pm 0.5$. This is consistent with the leading frequency dependence ($\psi_T \propto (Mf)^{5/3}$) of the tidal correction in Eq.~\eqref{eq:1pntaylorf2}. In addition, the free parameters $G$ and $H$ in this fit are highly correlated. We thus rewrite $E = G(Mf - Mf_0)^{5/3}$ and fit each waveform with the single parameter $G$. For fixed $\eta$ and $\chi_{\rm BH}$, $G$ is a roughly constant function of $\Lambda$, so we use the following form $G(\eta, \chi_{\rm BH}) = e^{g_0 + g_1 \eta + g_2 \chi_{\rm BH} + g_3 \eta \chi_{\rm BH} }$. The function $E$ can then be written
\begin{equation}
\label{eq:Efit}
E(Mf; \vec\theta) = e^{g_0 + g_1 \eta + g_2 \chi_{\rm BH} + g_3 \eta \chi_{\rm BH} }(Mf - Mf_0)^{5/3},
\end{equation}
where the best-fit parameters are $\{g_0, g_1, g_2, g_3\} = \{-1.9051, 15.564, -0.41109, 5.7044\}$, and as with the amplitude fit, this parametrization is well defined for all possible values of $\eta$, $\chi_{\rm BH}$, and $\Lambda$. We find that typical fractional errors in the fit for $G$ are $\sim 30\%$ for $\Lambda \gtrsim 500$, but can be larger for smaller values of $\Lambda$. As with the amplitude fit, the large fitting error for $\Lambda \lesssim 500$ is not significant because the systematic error resulting from the poor fit will still be less than the statistical error in $\Lambda$.

\subsection{Fit based on EOB BBH approximation}

Unlike the PhenomC waveform, EOB waveforms are not yet available for spins of $\chi_{\rm BH} = 0.75$, so we use only the 90 waveforms that have $-0.5 \le \chi_{\rm BH} \le 0.5$ when calibrating the fit. As discussed in Section~\ref{sec:spinEOB}, we Fourier transform the time-domain EOB waveform, then add the TaylorF2 tidal correction (Eq.~\eqref{eq:1pntaylorf2}) to the phase, and then generate a hybrid waveform. We then produce an analytic fit to these hybrid waveforms using the same procedure and matching parameters ($\Delta Mf_{\rm match} = 0.008$, $Mf_{\rm mid} = 0.022$, $Ms_i = Mf_i$, and $Ms_f = Ms_i + 0.001$) as with the PhenomC waveforms. 

For the amplitude fit we use for Eq.~\eqref{eq:Bfit}, $Mf_A = 0.01$ and obtain the following coefficients:\\ $\{b_0, b_1, b_2, c_0, c_1, c_2\}$\\$= \{-1424.2, 6423.4, 0.84203, -9.7628, 33.939, 1.0971\}$. \\ Typical errors in $C$ are about the same as for the PhenomC fit. 

For the phase fit we use for Eq.~\eqref{eq:Efit}, $Mf_\Phi = 0.02$ and obtain\\ 
$\{g_0, g_1, g_2, g_3\}  = \{-4.6339, 27.719, 10.268, -41.741\}$. \\Typical errors in $G$ are again about the same as for the PhenomC fit.

\section{Measurability of $\Lambda$}
\label{sec:7}
\subsection{Statistical error}

\begin{figure*}[!htb]
\begin{center}
\includegraphics[width=70mm]{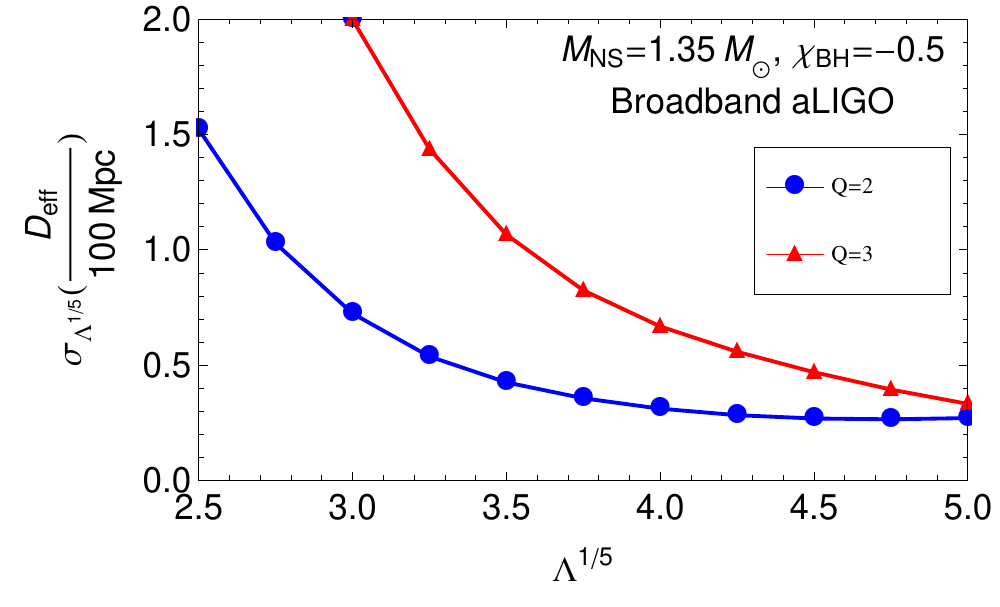}
\includegraphics[width=70mm]{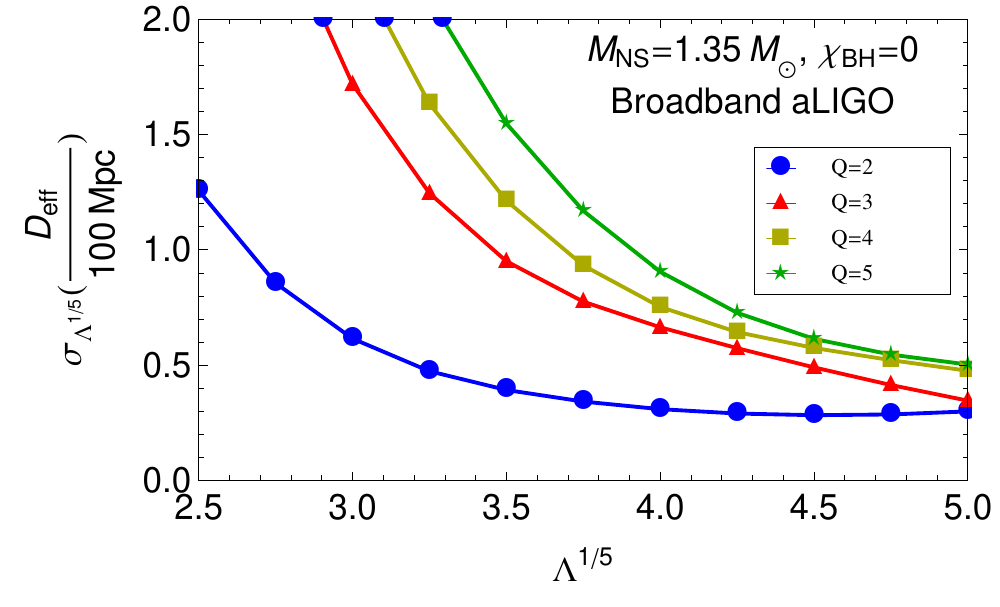}\\
\includegraphics[width=70mm]{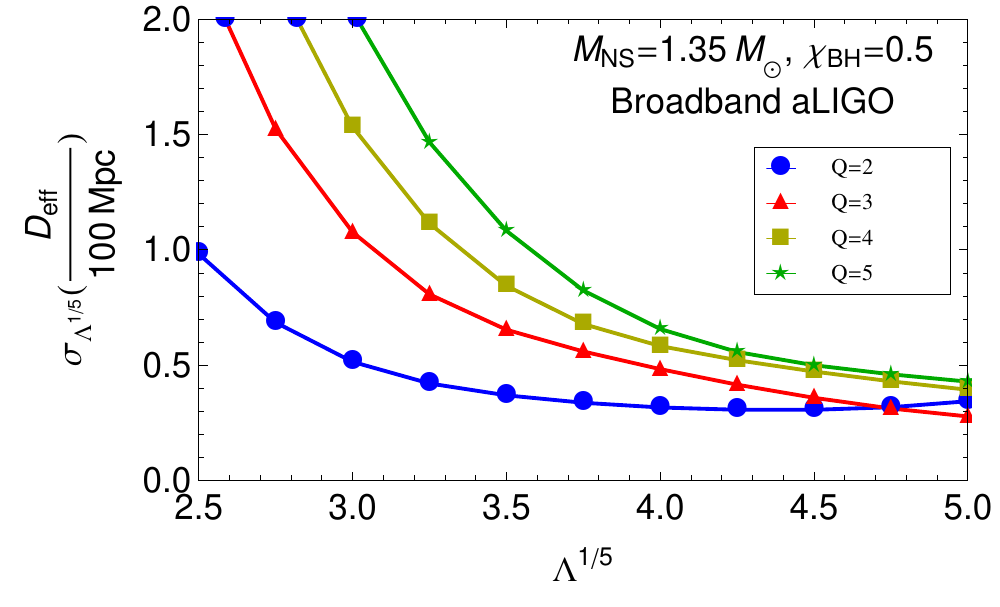}
\includegraphics[width=70mm]{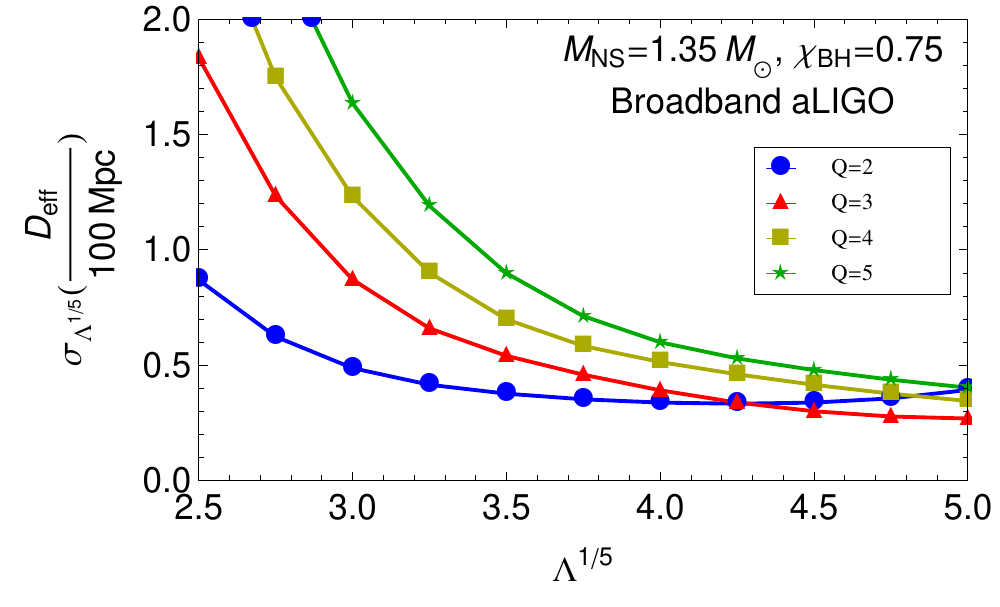}\\
\end{center}
\caption[Uncertainty in $\Lambda$ for broadband aLIGO]{ 
\label{fig:broaderror}
1-$\sigma$ error $\sigma_{\Lambda^{1/5}}$ for various values of the mass ratio, BH spin, and tidal deformability. NS mass is fixed at 1.35~$M_\odot$. The noise curve is for broadband aLIGO.
}
\end{figure*}

\begin{figure*}[!htb]
\begin{center}
\includegraphics[width=70mm]{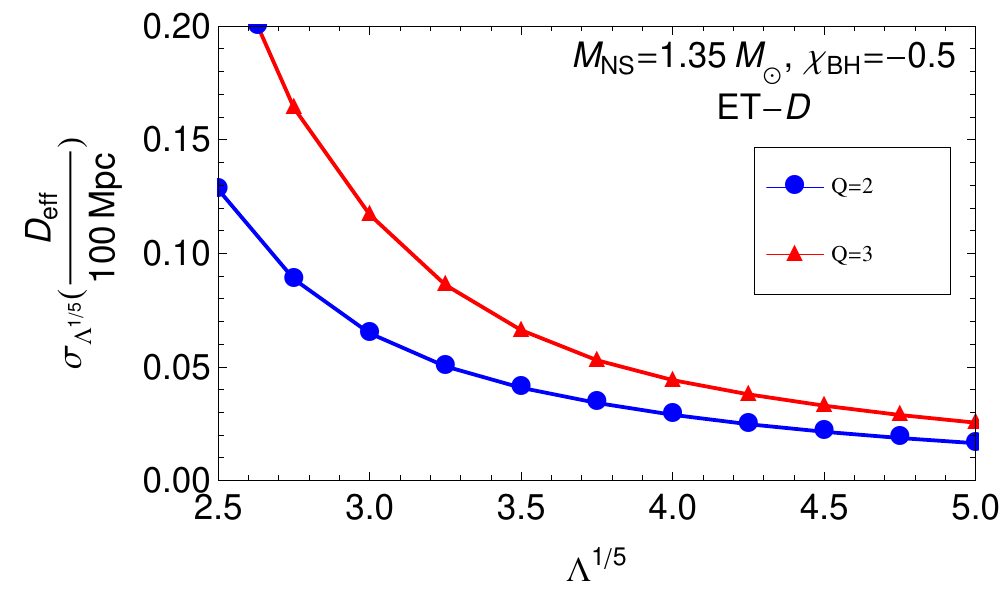}
\includegraphics[width=70mm]{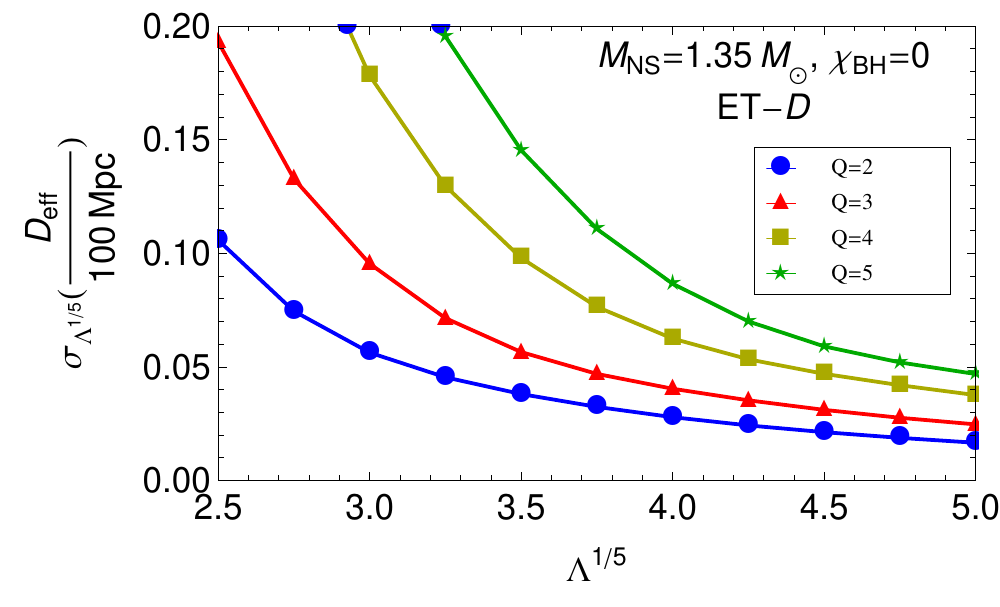}\\
\includegraphics[width=70mm]{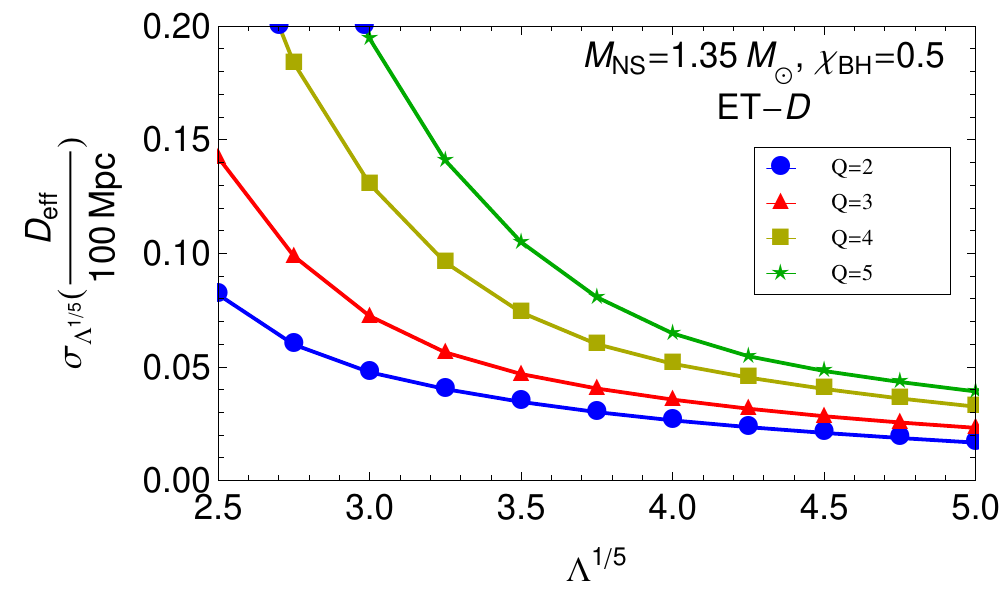}
\includegraphics[width=70mm]{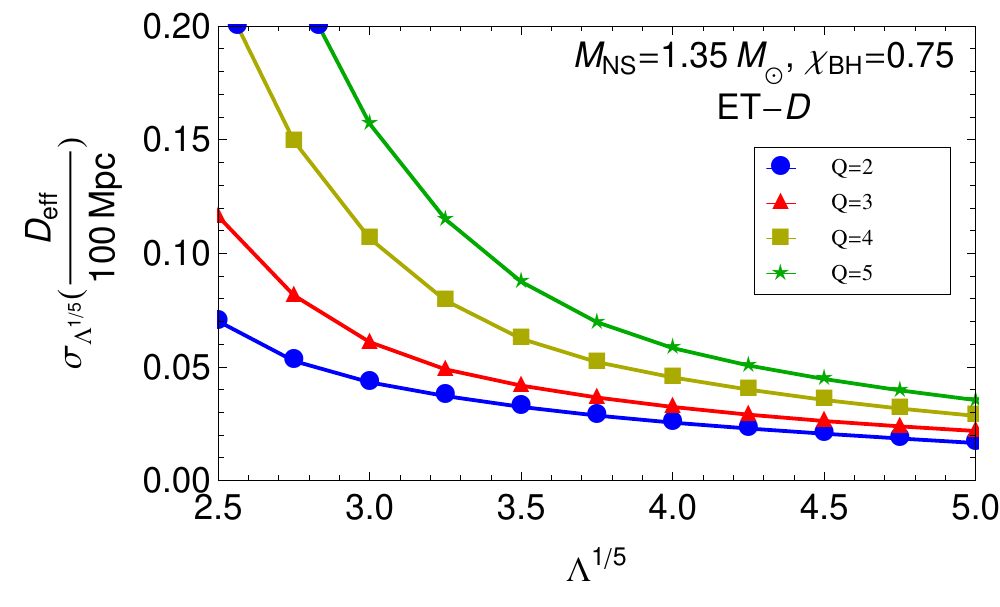}\\
\end{center}
\caption[Uncertainty in $\Lambda$ for ET-D]{ 
\label{fig:ETerror}
Same as Fig.~\ref{fig:broaderror}, but with the ET-D noise curve. Uncertainty $\sigma_{\Lambda^{1/5}}$ is an order of magnitude smaller.
}
\end{figure*}

Using the analytic BHNS waveform based on the PhenomC BBH waveform developed in the previous section, we can now evaluate the Fisher matrix for a single gravitational-wave detector using the complete set of waveform parameters \{$\ln D_{\rm eff}$, $f_1t_c$, $\phi_c$, $\ln\mathcal{M}$, $\ln\eta$, $\chi_{\rm BH}$, $\Lambda^{1/5}$\}, where $f_1$ is some fiducial frequency such as 1~Hz, and as in Paper~I we use $\Lambda^{1/5}$ because it is approximately proportional to the more familiar NS radius. We have calculated the 1-$\sigma$ uncertainty in $\Lambda^{1/5}$ for both the broadband aLIGO~\cite{LIGOnoise} and ET-D detector configurations~\cite{ETDnoise} shown in Fig.~\ref{fig:noisecurves}. Errors are shown in Figs.~\ref{fig:broaderror} and~\ref{fig:ETerror} for broadband aLIGO and ET-D respectively, and are scaled to an effective distance of 100~Mpc as was done in Paper~I. We note that the results here for the $Q = 2$ and 3 nonspinning waveforms are similar to those presented in Figs.~12 and~13 of Paper~I. This indicates that coherently adding the inspiral tidal interactions to the merger and ringdown, and considering correlations between $\Lambda$ and the other parameters roughly cancel each other.

There are several trends to notice in the uncertainty $\sigma_{\Lambda^{1/5}}$. In general, $\sigma_{\Lambda^{1/5}}$ increases with increasing mass ratio $Q$. This is not surprising since the inspiral tidal contribution $\psi_T$ (Eq.~\eqref{eq:1pntaylorf2}) to the waveform phase, which has a significant impact on the measurability of $\Lambda$ both before and after the inspiral-merger transition, is a decreasing function of the mass ratio. In addition, the amount of tidal disruption before the plunge, as well as its imprint on the waveform, decreases as the mass ratio increases. However, there are two competing effects that help to minimize the increase in uncertainty $\sigma_{\Lambda^{1/5}}$ as $Q$ increases. First, the amplitude during the inspiral which to Newtonian order scales as $|\tilde h(f)| \propto \mathcal{M}^{5/6}f^{-7/6}/D_{\rm eff}$ increases as the mass ratio increases for a fixed NS mass. Second, for higher mass ratios, the EOS dependent merger dynamics occur at lower frequencies, closer to the minimum of the noise PSD (Fig.~\ref{fig:noisecurves}).

On the other hand, $\sigma_{\Lambda^{1/5}}$ decreases with increasing black hole spin $\chi_{\rm BH}$. This effect can be understood from Fig.~\ref{fig:hybridvarychi} where the amplitude and phase difference between a BHNS waveform and BBH waveform with the same parameters increases as the BH spin increases, and the amplitude cutoff occurs at a lower frequency where the detector is more sensitive. Physically, we expect the EOS dependence to be greater for higher spins because the BH ISCO decreases with spin, allowing the NS to become more tidally distorted before passing through the ISCO and plunging into the black hole. In addition, for high spins, the orbital decay will take longer because the system must radiate away sufficient angular momentum for the final Kerr black hole to have spin parameter $\chi^{\rm final}_{\rm BH} < 1$. As a result, the waveform has more chance to deviate from a BBH waveform towards the end of the inspiral and into merger.

We also note that in general, $\sigma_{\Lambda^{1/5}}$ decreases as a function of $\Lambda^{1/5}$. This occurs because the departure from BBH behavior, given by Eqs.~\eqref{eq:r} and~\eqref{eq:deltaphi}, is a strongly increasing functions of $\Lambda^{1/5}$. However, as seen in Fig.~\ref{fig:broaderror} for broadband aLIGO, we find that for $Q=2$ and $\chi_{\rm BH}\gtrsim 0.5$, the error $\sigma_{\Lambda^{1/5}}$ begins to increase again for large $\Lambda^{1/5}$. Because the Fisher matrix element $\Gamma_{\Lambda^{1/5} \Lambda^{1/5}}$ is a monotonically increasing function of $\Lambda^{1/5}$, the increase in $\sigma_{\Lambda^{1/5}}$ is, therefore, due to an increase in the covariance with the other parameters for this particular set of parameters and noise curve. Finally, we find that although the uncertainty $\sigma_{\Lambda^{1/5}}$ for the parameter values ($Q=2$, $\chi_{\rm BH}=0$, $M_{\rm NS}=1.35M_\odot$), ($Q=3$, $\chi_{\rm BH}=0.5$, $M_{\rm NS}=1.35M_\odot$), and ($Q=5$, $\chi_{\rm BH}=0.75$, $M_{\rm NS}=1.35M_\odot$) only varied by $\sim 50\%$ in Fig.~\ref{fig:ellipseETIMR}, they vary by $\sim 100\%$ in Figs.~\ref{fig:broaderror} and~\ref{fig:ETerror}. This again results from accounting for the covariance with the other parameters.

\subsection{Systematic error} 

The analytic BHNS waveform developed in Section~\ref{sec:phenomBHNS} does not exactly match the hybrid waveforms that it is calibrated against. This leads to a bias in estimating the BHNS parameters. In this subsection, we estimate the systematic error that results from using our analytic waveform instead of the hybrid waveforms as templates, then we use it as a criteria to determine the quality of the analytic fit.

In section~\ref{sec:estimation}, we gave an approximate expression for the systematic error
\begin{equation}
\label{eq:systerror}
\Delta\theta_{\rm syst}^i \approx - \Gamma^{-1}_{ij} (\delta h(\vec\theta_A), \partial_j h_A(\vec\theta_A)),
\end{equation}
where $\delta h(t; \vec\theta) = h_A(t; \vec\theta) - h_E(t; \vec\theta)$ is the difference between the approximate fit and exact waveform. Here, we refer to the hybrid BHNS waveform as the exact waveform $h_E(t; \vec\theta)$, the analytic fit as the approximate waveform $h_A(t; \vec\theta)$, and $\vec\theta_A$ is the best estimate of the true parameters using the approximate fit as the template. The waveform derivatives and Fisher matrix $\Gamma_{ij}$ are then calculated from the analytic waveform fit as was done for the statistical error.

An analytic waveform is useful for parameter estimation when the systematic error is a small fraction of the statistical error for a given gravitational wave detector. We will use as our criteria for a sufficiently accurate fit to the numerical waveform the requirement that the systematic error in each parameter be less than the statistical error for an optimally oriented BHNS system observed at 100~Mpc. For larger distances or a less sensitive detector, the systematic error will be a smaller fraction of the total error. For a waveform with the form $h(t)=\frac{1}{D_{\rm eff}}g(t)$ and a PSD with overall amplitude factor $A_n$ such that $S_n(f) = A_n^2 R_n(f)$, the statistical errors $\sigma_i$ scale as $\sigma_i \propto A_n D_{\rm eff}$. On the other hand, as can be found from Eq.~\eqref{eq:systerror}, the systematic error is independent of both $A_n$ and $D_{\rm eff}$. The ratio of systematic to statistical error is therefore
\begin{equation}
\frac{\Delta\theta_{\rm syst}^i}{\sigma_i} \propto \frac{1}{A_n D_{\rm eff}}.
\end{equation}
Because the PSD for broadband aLIGO and ET-D have roughly the same shape, but differ in amplitude $A_n$ by a factor of $\sim 10$, we expect the systematic to statistical error ratio to differ by about a factor of 10, and therefore the systematic error will be far more important for ET. 

We find, for the BHNS fit based on the PhenomC waveform, the ratio of systematic error to statistical error in $\Lambda^{1/5}$ for broadband aLIGO for a binary at $D_{\rm eff} = 100$~Mpc is $|\Delta\Lambda^{1/5}_{\rm syst}| / \sigma_{\Lambda^{1/5}} \sim 0.15 \pm 0.15$ with a maximum value of 0.65 for the 134 waveforms listed in Table~\ref{tab:simulations}. However, the systematic error is strongly biased by the $Q = 5$ waveforms for which the PhenomC waveform has not been calibrated to BBH simulations (Table~\ref{tab:simulations}). For the 101 simulations with $Q \le 4$, we find $|\Delta\Lambda^{1/5}_{\rm syst}| / \sigma_{\Lambda^{1/5}} \sim 0.09 \pm 0.08$ with a maximum value of 0.51, and we also note that this ratio is $>0.2$ only when $Q \ge 4$.

For ET-D, the systematic error is roughly the same as for broadband aLIGO, and the ratio of systematic to statistical error is about an order of magnitude larger as expected. Specifically, at $D_{\rm eff} = 100$~Mpc, $|\Delta\Lambda^{1/5}_{\rm syst}| / \sigma_{\Lambda^{1/5}} \sim 1.3 \pm 1.2$ with a maximum value of 6.3 for the 134 waveforms. Only including the 101 waveforms where $Q \le 4$, $|\Delta\Lambda^{1/5}_{\rm syst}| / \sigma_{\Lambda^{1/5}} \sim 1.1 \pm 0.9$ with a maximum value of 5.3, and this ratio is $>2.5$ only when $Q \ge 4$. We therefore conclude that the BHNS fit is sufficient for aLIGO. However when the effective distance $D_{\rm eff}$ is less than a few hundred~Mpc, the systematic error will become comparable to the statistical error in some cases for ET-D.

We also calculate the systematic error for the fit based on the EOB waveform. Unfortunately, because the EOB waveform is time-domain and the waveform enters the detector band at around 10~Hz for aLIGO and 1~Hz for ET, very long waveforms are needed to evaluate the statistical and systematic error. This is possible, but time consuming, so we use a more crude estimate of the errors. Eqs.~\eqref{eq:sigma} and~\eqref{eq:systerror} can be approximated by ignoring the covariance between $\Lambda$ and the other parameters:
\begin{align}
\sigma_{\Lambda^{1/5}} &\approx \left[(\partial_{\Lambda^{1/5}} h_A(\vec\theta_A), \partial_{\Lambda^{1/5}} h_A(\vec\theta_A))|_{Mf=0.01}^{Mf=\infty}\right]^{-1/2}, \\
(\Delta\Lambda^{1/5})_{\rm syst} &\approx - \frac{(\delta h(\vec\theta_A), \partial_{\Lambda^{1/5}} h_A(\vec\theta_A))|_{Mf=0.01}^{Mf=\infty}}{(\partial_{\Lambda^{1/5}} h_A(\vec\theta_A), \partial_{\Lambda^{1/5}} h_A(\vec\theta_A))|_{Mf=0.01}^{Mf=\infty}},
\end{align}
where the integral in the inner product $(\cdot | \cdot)|_{Mf=0.01}^{Mf=\infty}$ is evaluated from a frequency of $Mf=0.01$ to $\infty$. We also note that the hybrid waveform $h_E$ and analytic fit $h_A$ are identical below $Mf = 0.01$, so $\delta h=0$ below this frequency.

We have evaluated these quantities for the BHNS fits based on the PhenomC and EOB waveforms. The statistical errors are the same to approximately $\pm 10\%$. The systematic error for the EOB fit, however, is typically $\sim 2$ times larger than for the PhenomC fit. This is somewhat surprising given that the EOB waveform is a much better approximation to BHNS waveforms with the smallest values of $\Lambda$. However, we note that most of the adjustable parameters, such as the matching window, as well as the form of the functions $r$ and $\Delta\Phi$ (Eqs.~\eqref{eq:r}--\eqref{eq:Efit}) were optimized for hybrids based on the PhenomC waveform. We therefore do not claim that the PhenomC waveform in general produces a better BHNS analytic fit.

\section{Discussion}
\label{sec:8}
\subsection{Summary}

We have examined the ability of gravitational wave detectors to extract information about the EOS from observations of BHNS coalescence for black holes with aligned spin. In Paper~I, we found that the EOS parameter that is best measured during the merger and ringdown, for systems with nonspinning black holes, is consistent with the tidal deformability $\Lambda$. We have now found that this is also true for systems with aligned black hole spins. Furthermore, coherently joining the tidally corrected inspiral, which analytically depends on $\Lambda$, to the merger and ringdown dramatically improves the alignment of the error ellipses with $\Lambda$ in Fig.~\ref{fig:ellipseETIMR} as well as the measurability of $\Lambda$ by up to a factor of $\sim 3$ in some cases over just the merger and ringdown. 

In order to examine the correlations between $\Lambda$ and the other parameters, we constructed an analytic IMR waveform based on the frequency-domain, aligned-spin PhenomC BBH waveform model~\cite{Santamaria2010} as well as the time-domain EOB waveform model~\cite{TaracchiniPanBuonanno2012}, and we calibrated this waveform model to our hybridized numerical waveforms. Although $\Lambda$ does correlate with the other parameters, the correlations are not nearly as strong as correlations between the other parameters. Overall, the correlations reduce the measurability of $\Lambda$ by approximately a factor of 3. The above two effects roughly cancel out and the results for $\sigma_{\Lambda^{1/5}}$ are therefore similar to the results presented in Paper~I which neglected these two effects. 

In addition, we examined the agreement between the hybrid BHNS waveforms and our analytic waveforms. The agreement is good enough that systematic errors will be smaller than statistical errors for aLIGO. However, for ET, systematic errors will matter if BHNS systems are observed with effective distances of less than a few hundred Mpc.

\subsection{Future Work}

There is currently enough uncertainty in the modeling of BBH systems and inspiral tidal interactions, as well as in BHNS simulations and the hybridization procedure, that the analytic BHNS waveforms presented here can only be considered preliminary. Additional systematic errors exist. The waveform model has the ability to incorporate improvements in the inspiral point-particle and tidal interactions. However, incorporating EOS dependent corrections into the merger-ringdown waveform requires an accurate understanding of the late dynamics for both BBH and BHNS systems. This analytic waveform fit will therefore need to be re-calibrated when improvements are made to the late dynamics. On the numerical side for BHNS systems, key improvements beyond the standard issue of convergence, would be to minimize eccentricity in the numerical waveforms and to increase the number of orbits before merger so that the numerical waveform can be matched to the inspiral waveform at lower frequencies where errors in the analytic point-particle and tidal interactions are smaller. On the analytic side, we are restricted to analyzing correlations between parameters for systems with small mass ratios and moderate black hole spins because IMR BBH models have not yet been calibrated to BBH simulations with larger mass ratios and spins.

The analytic waveform models presented here, and in particular the one based on the frequency-domain PhenomC waveform,  can be, without too much difficulty, incorporated into Markov Chain Monte Carlo and Nested Sampling algorithms used for Bayesian parameter estimation for networks of gravitational-wave detectors. A full Bayesian analysis will then make it possible to assess the true nature of the statistical and systematic errors beyond the Fisher matrix approximation, by injecting hybrid BHNS waveforms into detector noise and attempting to recover their parameters with the analytic waveform template. 

\acknowledgments

We thank Lucia Santamaria for providing code for calculating PhenomC waveforms, Alessandra Buonanno and Andrea Taracchini for generating spinning EOB waveforms, Richard O'Shaughnessy for helpful discussions related to the phenomenological fits, and Marc Favata for helpful discussions related to the measurability of tidal parameters. This work was supported by NSF Grants PHY-1001515, PHY-0970074, and PHY11-25915. The work of MS is supported by Grant-in-Aid for Scientific Research (No. 21340051, 24740163), Grant-in-Aid for Scientific Research on Innovative Area (No. 20105004), and HPCI Strategic Program of Japanese MEXT. The work of KK is supported by Grant-in-Aid for Scientific Research (No. 21684014). Part of this work was done while BL was at KITP.

\bibliography{spinning}

\end{document}